\documentclass[preprint,12pt,letterpaper]{aastex}

\shorttitle{Big Blue Bump}
\shortauthors{Atlee \& Mathur}

\begin{document}

\title{GALEX Measurements of the Big Blue Bump in Soft X-ray Selected AGN}

\author{David W. Atlee \and Smita Mathur}
\affil{Department of Astronomy, The Ohio State University}
\email{atlee@astronomy.ohio-state.edu}

\begin{abstract}
We study the UV properties of Type I AGN from the ROSAT All-Sky Survey that
have been selected to show unusually soft X-ray continua.  We
examine a sample of 54 Seyfert 1 galaxies 
with detections in both Near-UV and Far-UV bands of the Galaxy Evolution
Explorer (GALEX) satellite.
Our sample is systematically fainter in the UV than galaxies studied 
in similar work by previous authors.
We look for correlations between their UV and X-ray properties as well
as correlations of these properties with either black hole
mass or Eddington ratio.  The shape of the Big Blue Bump 
(BBB) in the GALEX regime does not appear to correlate with
its strength relative to the power law continuum, which conflicts
with results reported by previous authors.  
The strength of the BBB is correlated with the shape of the
X-ray continuum, in agreement with previous work, but the slope of the
correlation is different than previously reported.
The properties of the accretion
disks of Type I AGN in the GALEX regime
are relatively independent of black hole mass
and Eddington ratio.  We compare our measurements to the
predictions of alternative theories for the origin of the soft excess, but
we are unable to distinguish between Comptonization of BBB photons
by a hot plasma and absorption in relativistic winds as the most likely
origins for the soft X-ray excess.
\end{abstract}

\keywords{galaxies: Seyfert, ultraviolet: general, X-ray: general}

\section{Introduction}\label{secIntro}
The soft X-ray excess is a contribution to the 0.2-2 keV flux 
in some Type I AGN beyond that predicted by 
extrapolating the hard X-ray power-law.  It
was first reported by \citet{arna85}, who suggested that it was caused by
thermal emission from the hot inner portion of the AGN accretion disk.  
Early observational work by \citet{turn89} using EXOSAT and subsequently
by Walter \& Fink (1993; WF93) using ROSAT further explored the properties
of the soft excess in an attempt to conclusively determine its origin.
\citet{turn89} found their results to be consistent with the
soft excess arising from the high-energy tail of
the thermal accretion disk emission for AGN, but WF93 found
that their measurements using a combination of X-ray
measurements from ROSAT and Ginga and 
UV fluxes from the International Ultraviolet Explorer (IUE)
were inconsistent with simple thick or thin accretion disk models.
They also discovered that the ROSAT spectral indices ($\alpha_{\rm x}$)
of AGN with soft excesses were strongly correlated with the 
strength of the excess.
The results of WF93 were later verified by \citet{walt94} using 
simultaneous IUE and ROSAT observations.

Several alternative theories for the origin of the soft excess have 
since been proposed. The current models tend to favor
either reprocessing of thermal disk emission via Compton scattering in
thermal plasmas (e.g. \citealt{kawa01}, \citealt{nied06}) or 
relativistically broadened absorption (e.g. \citealt{schu07}).
Models invoking atomic processes
to generate the soft excess were originally proposed by \citet{gier04}, who
noted that the soft excess shows very consistent ``temperature''
across AGN with a wide variety of black hole masses.
\citet{schu06} recently proposed an alternative picture
to the usual wind model for atomic origins.  Their ``failed''
wind model does not require the massive outflow from the accretion
disk usually required to make an atomic origin viable.

Several additional theories have also been proposed, including Compton
reflection of hard X-ray photons by the dense, low-ionization gas
in the accretion disk, resulting in an emergent spectrum that
is very steep in the soft X-ray regime (\citealt{ross93}; \citealt{sobo07};
\citealt{done07}).
Alternatively, hard X-ray photons could be absorbed by the 
disk instead of being reflected, and the absorbed energy
would be re-emitted as soft 
X-ray photons with spectral indices that depend on the properties
of the disk \citep{roza02}.
Another popular class of model is
the slim disk model, originally proposed by \citet{much82}, in which
super-Eddington accretion causes changes in the properties of the standard
thin disk, causing it to become geometrically thick and optically 
thin in its inner region and emit high-energy photons \citep{chen04}.
For more recent theoretical treatments of slim disks, see e.g.
\citet{hein06} and \citet{hein07}.

After WF93, much of the observational work on the
soft excess focused on the X-ray
properties at the expense of the UV.  The seminal
paper from \citet{boll96}, which first reported the different distributions
of $\Gamma_{\rm x}$ seen in Narrow-Line Seyfert 1 galaxies (NLS1s)
and normal Seyfert 1 galaxies (BLS1s) is one example.  However, it is
often difficult to distinguish between the various competing models
based only on goodness of fit to X-ray spectra (e.g. \citealt{sobo07};
\citealt{piro97}),
and each of the competing models has its drawbacks.  Comptonization
models require nearly constant temperatures, and absorption models tend
to produce sharper absorption lines than desirable, for example.

In Grupe et al. (1998, 2004; G98, G04), large samples of Seyfert 1 galaxies 
with strong soft excesses were drawn from the full set
of optically-identified ROSAT All-Sky Survey (RASS) sources.
Some of their conclusions were similar to those of
WF93, but in neither paper did the authors examine the
UV fluxes of their sample.  Since the publication of WF93, a number
of papers have examined
the UV properties of a handful of soft excess AGN (\citealt{puch95a};
\citealt{puch95b}) and NLS1 galaxies (\citealt{kura00}; \citealt{leig04}).
Other work has focused on large samples of AGN with data from
multiple wavelength regimes (e.g. \citealt{stra05}, \citealt{main07}
\citealt{kell08}),
but there have been no attempts to study the UV properties of a 
moderately large, uniformly-selected sample of soft excess AGN.

The samples of G98 and G04 were selected to show unusually soft
ROSAT spectra, so these AGN all show significant soft
excesses.  The samples were also selected uniformly in their X-ray
properties, and they contain reasonably large numbers of objects.
We use Galaxy Evolution Explorer (GALEX) fluxes to study 
the UV properties of these objects, determining the shape and 
relative strength of the Big Blue Bump (BBB).  By measuring the
BBB directly and relating its properties to the soft excess, we
can provide additional constraints on the physical mechanism responsible
for generating the soft X-ray excess.  In \S\ref{secObservations} we
discuss the observations we collected and the extraction of the necessary
parameters.  In \S\ref{secAnalysis} we discuss the analysis
we perform on the extracted parameters, and in \S\ref{secModels}
we compare our results with the predictions of theories for the 
origin of the soft excess.

Luminosities in this paper are calculated using $H_{0}=72\ {\rm km s^{-1}}$,
$\Omega_{m}=0.27$ and $\Omega_{\Lambda}=0.73$.

\section{AGN Sample}\label{secObservations}
Grupe et al. (1998, 2004) selected samples of soft,
X-ray bright AGN at high Galactic latitude from the RASS.
They required that their objects all have ROSAT hardness ratios (HR)
less than zero, yielding a sample of AGN with relatively strong 
soft excesses.  
We acquired GALEX Release 3 (GR3)\footnote{http://galex.stsci.edu/GR3}
images of the G98 and G04 AGN wherever possible.  Measuring
the Near-UV ($NUV$, $\lambda_{eff}=2271$\AA) and Far-UV 
($FUV$, $\lambda_{eff}=1528$\AA) fluxes from the GR3 images,
we constructed a sample of 54 AGN with measurements of both the
UV and soft X-ray fluxes.  These AGN are listed in Table \ref{tabGalaxies}.

We extracted X-ray count rates and spectral indices for the
AGN in our sample from Grupe et al. (2001, 2004) wherever possible, 
and from Grupe et al.
(1998, 1999) otherwise.  Following the convention in these papers, 
we list X-ray spectral indices in energy units, i.e.
$F_{\nu} \propto \nu^{-\alpha_{\rm x}}$, where $\alpha_{\rm x}$ is measured
in the ROSAT band (0.2-2.0 keV).  This differs from the
convention of WF93, who list photon indices ($\Gamma_{\rm x}=\alpha_{\rm x}+1$,
where $N_{\nu}\propto\nu^{-\Gamma_{\rm x}}$).  We have accounted
for this difference
when comparing with their results.
The spectral indices for galaxies in the Grupe et al.
catalogs are higher than for the average AGN, as expected for a sample
of soft excess AGNs, but our $\langle\alpha_{\rm x}\rangle$ is
even higher than the average of the WF93 sample (2.1 compared to 1.5 in
WF93).  This shift 
in $\langle\alpha_{\rm x}\rangle$ is due
primarily to the inclusion of a large Narrow-Line
Seyfert 1 (NLS1) subsample, as NLS1s are known to exhibit softer
X-ray continua (larger $\alpha_{\rm x}$; \citealt{boll96}).

Our AGN sample has only three objects (NGC 4593, Mrk 142 \& 478) in common with
the WF93 sample, despite covering roughly the same ranges in
redshift and X-ray flux.  This is due to selection effects,
as WF93 required that their objects have UV (International 
Ultraviolet Explorer; IUE), 5 GHz radio continuum and hard X-ray (Ginga)
flux measurements in addition to their ROSAT fluxes.
We supplemented the information in the Grupe et al. catalogs with 
IRAS 25$\mu m$ fluxes from the NASA Extragalactic Database
(NED\footnote{http://nedwww.ipac.caltech.edu}, 
various authors).  We found 25$\mu m$ fluxes for only 17 of the 54
objects in our sample, so we use the IRAS fluxes only to verify that our
primary strength indicator for the BBB is unbiased with 
respect to the strengths reported by WF93 (see Section \ref{secObservables}).

The majority of the GALEX images available for 
our sample come from the GALEX All-Sky Imaging Survey (AIS), which
have $\sim$100 s exposure per field.  Two objects
(Mrk 1048 and PG 1244+026) also had deeper Medium Imaging Survey 
exposures, and Markarian 1048 had another, still deeper, exposure from 
the Guest Investigator program.  QSO 0056-36 also had a second AIS exposure.
The GALEX exposure times associated with each object are listed in Table
\ref{tabGalaxies}, along with several other important parameters 
including UV fluxes, X-ray count rates and
H$\beta$ line widths.
We required that all objects in our sample have detections in both the
$NUV$ and $FUV$, but this restriction did not result in excluding any objects
from our sample.

We divided our AGN into Narrow-Line Seyfert 1 (NLS1) and normal 
Seyfert 1 (BLS1) classes based on the width of the
${\rm H}\beta$ emission line.  All objects with 
${\rm FWHM(H}\beta)<2000\ {\rm km\ s^{-1}}$ were classified as NLS1s.
We used the line widths listed in \citet{grup04a} wherever 
possible and widths from \citet{grup99} otherwise, classifying
29 of our 54 AGN as NLS1s.
Markarian 734, which is listed only in the \citet{grup99} catalog,
has FWHM(H$\beta$) and $\alpha_{\rm x}$ similar to four NLS1s that would
be classified as BLS1s based on their \citet{grup99} line widths
(see Table \ref{tabReclassified}).  
This suggests that the classification of Mrk 734 as a BLS1 may be
erroneous.  Removing it from the sample has no significant effect on our
conclusions.

\subsection{Flux Extraction}
We measured GALEX $FUV$ and $NUV$ count rates within $18''$ photometric
apertures for each AGN in our sample.  We converted the measured count rates
to magnitudes using the GALEX photometric zero-points of \citet{morr05},
\begin{equation}
m_{FUV}=-2.5\log(C_{FUV})+18.82
\end{equation}
\begin{equation}
m_{NUV}=-2.5\log(C_{NUV})+20.08
\end{equation}
where $C_{x}$ is the count rate in bandpass $x$.  To compute
the Galactic extinction corrections, we averaged the
reddening law of \citet{card89} across the $FUV$ and $NUV$ effective
area curves,
\begin{equation}\label{eqExtinction}
R_{\rm X} = \frac{\int^{\lambda_{2}}_{\lambda_{1}} R(\lambda)T(\lambda)d\lambda}
{\int^{\lambda_{2}}_{\lambda_{1}} T(\lambda)d\lambda},
\end{equation}
where $R(\lambda)$ is the \citet{card89} R-value at wavelength $\lambda$, 
and $T(\lambda)$ is the filter bandpass. We found ${\rm R}_{FUV}=8.24$ 
and ${\rm R}_{NUV}=8.10$. The R-values and
${\rm E}\left(B-V\right)$ color excesses for each line of sight \citep{schl98}
were used to compute total extinction and correct the measured 
fluxes for each object.  We used the AB-magnitude relation to convert
the dereddened magnitudes to UV flux densities, and we 
used PIMMS to solve for $F_{\nu}({\rm 2\ keV})$
for our AGN using the X-ray count rates and spectral indices
recorded in the Grupe catalogs.  Grupe et al. fixed column 
densities at the Galactic
value unless $N_{H,fit}>N_{H,gal}+2\times10^{20}\ {\rm cm^{-2}}$, 
in which case the fitted value was used.
Determining $F_{\nu}$ in this way assumes that $\alpha_{\rm x}$
provides a good description
of the X-ray spectrum across the entire ROSAT energy range.
This assumption is not perfect and is likely to do worse in objects
with steeper spectra, so some results may be biased.
However, the number of objects with extremely steep spectra is limited
(4 objects with $\alpha_{\rm x}>3$), and the uncertainties on these
indices are relatively large, so any biases in the measurements are
covered by the error budget.

\subsection{Black Hole Masses}\label{secBlackHole}
We use black hole masses and Eddington ratios ($L/L_{\rm edd}$) 
from \citet{grup04c}, which include 30 of our 54 AGNs.  These black hole
masses were determined using the \citet{kasp00} relations, which
was calibrated empirically using reverberation
mapped AGNs.  The calibration sample included several AGN
with luminosities similar to the AGN in our sample, so
the Kaspi relation
should yield reasonably robust black hole masses.
\citet{bent06} reported a different scaling relation
with a power law index of $0.52\pm0.04$, which is in agreement with the value
expected from theory.  If we use the Bentz relation instead of the
Kaspi relation, we find significantly larger black hole masses
than computed by \citet{grup04c}, but this is not unexpected since
the stellar continuum has not been subtracted from $L_{5100\ \AA}$
for our AGN sample, resulting in an over-estimate of the radius of
the Broad Line Region (BLR).
The alternative masses have a significant impact on some
of the measured correlations, which can be seen by comparing the
correlation coefficients in Tables \ref{tabBentz} and 
\ref{tabPhysicalCorrelation}.  Unsurprisingly, using the alternative
$L_{\lambda}(5100\ {\rm \AA})$--$R_{\rm BLR}$ relation has the largest impact
on correlations with $L_{\lambda}(5100\ {\rm \AA})$; the other changes are
rarely significant.  Because we
have no information on the host galaxies of our AGN,
we have elected to use the empirically calibrated Kaspi
relation.  This will add scatter to our inferred black hole
masses, but based on the few qualitative differences between 
Tables \ref{tabBentz} and \ref{tabPhysicalCorrelation} we
infer that the potential to introduce 
or hide correlations is limited.

We might also consider the impact of radiation pressure
on the derived black hole masses, as described by \citet{marc08}.
Qualitatively, the impact of radiation pressure should result in higher
masses for systems radiating closer to their Eddington rate, and
\citet{marc08} found that, after applying this correction, the Eddington
ratios of NLS1 systems are less extreme than result from applying the
Bentz and Kaspi relations.  Computing black hole masses using the Marconi
relation rather than the Kaspi relation yields masses that are, on average,
0.2 dex larger.  The new masses are correlated with the old masses
with $r_{s}=0.69$.
However, Marconi et al. failed to account for
the lower Eddington ratios and reduced radiation pressure implied by
the adjusted black hole masses.  As a result, their results overestimate
black hole masses in systems where the corrections for radiation pressure
are significant, and the difference between the ``true'' black hole masses
and the results calculated using the Kaspi relation will be less than
the 0.2 dex implied by applying the Marconi relation.

\citet{netz09} found that black hole masses determined using the Marconi 
relation are distributed differently from black hole masses in type 2 AGNs,
suggesting that radiation pressure is not important in nearby AGNs.  However, 
\citet{marc09} in turn suggested that the differences found by \citet{netz09}
can be attributed to scatter in the underlying scaling relations rather than
a lack of radiation pressure support in the BLR.  Nevertheless,
\citet{marc09} and \citet{netz09} agree that the \citet{marc08}
relation is unable to successfully reproduce the ``true,'' underlying
mass distribution, indicating that more work is needed.
For the rest of this paper,
we consider masses resulting from the Kaspi relation with the caveat
that the systematic uncertainties associated with the alternative methods
for calculating black hole mass must also be considered.

The masses resulting from applying the Kaspi relation are more properly
called virial products, which differ from the true black hole
mass by a geometric factor $f$.  There is significant debate in the
literature on the proper value of this constant.
Using the dispersion of the H$\beta$\ emission line to 
measure the virial products, \citet{onke04} found that a statistical 
correction of $f=5.5$ was required to bring their virial masses
into agreement with black hole masses predicted by the
${\rm M_{BH}}-\sigma_{*}$ relation.  
By contrast, \citet{wats07} found a correction of $f=2.2$ for
AGN in the \citet{grup04a} sample using the line dispersion or $f=0.55$
using FWHM.  The latter value disagrees with the results of \citet{kasp00},
who found $f=0.75$.  \citet{wats07} also
found that there is a systematic difference between the geometric corrections
required to bring the BLS1 and NLS1 samples into agreement with the
$M_{BH}-\sigma_{*}$ relation.  Given this disagreement,
we choose not to apply a geometric
factor and simply use the virial products.
As a result, the absolute masses and Eddington ratios
we use are incorrect, but there will be 
little effect on the measured correlations as long as $f$
is a constant.  If the geometric corrections required by NLS1s and
BLS1s do indeed differ, the impact of using virial products instead
of actual black hole masses might be significant.

\section{Correlation Analysis}\label{secAnalysis}
In this section, 
we examine the differences in several measurable parameters between 
NLS1s and BLS1s.  We also study the relationships between
the parameters themselves.  We examine several
observables, including flux ratios and $\alpha_{\rm x}$, as well as the
physical characteristics ($M_{BH}$, $L/L_{\rm edd}$ and $L$)
that determine the properties of each AGN.
\citet{walt93} found that the strength of the soft X-ray excess,
measured using hard X-ray fluxes from Ginga and soft X-ray fluxes
from ROSAT, correlates well with the ROSAT spectral index
(WF93, Fig 7).  Based on this result, they used $\Gamma_{\rm x}$ as a
proxy for the strength of the soft excess.  We take a similar approach,
using $\alpha_{\rm x}$ instead of $\Gamma_{\rm x}$, to examine the
relationships between the soft X-ray excess and the UV
properties of the AGN in our sample.  However, it is important to note
that when we discuss the ``soft X-ray excess'' below, we actually mean
the shape of the soft X-ray continuum.

\subsection{Observables}\label{secObservables}
\begin{figure}
\epsscale{1.0}
\plotone{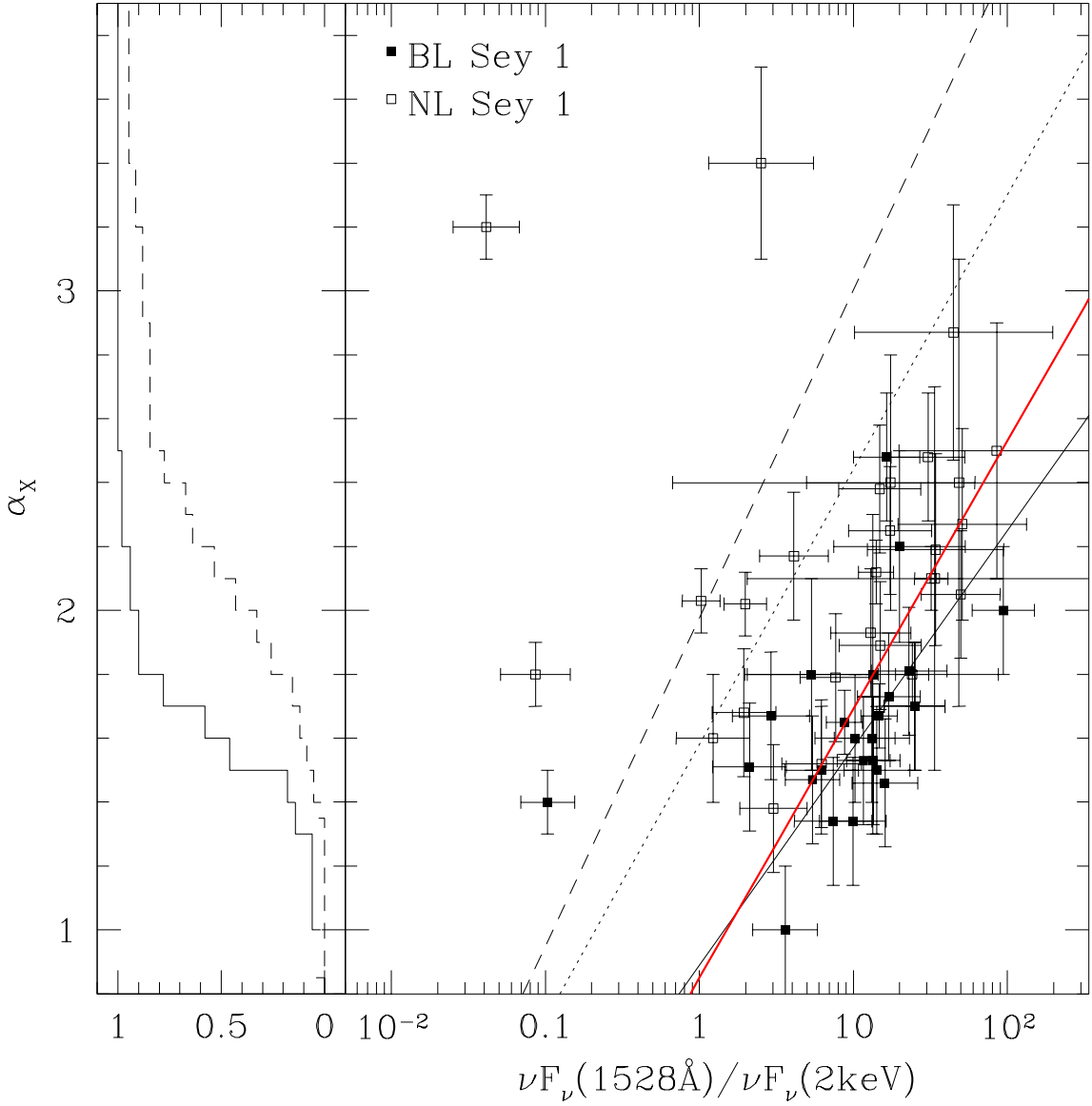}
\caption{Comparison of two indicators for the strength of the soft X-ray
excess.  UV flux density is measured at the effective
wavelength of the GALEX $FUV$ band using the AB-magnitude relation.
The {\it red} line is the best fit to our AGNs 
$\biggl(\alpha_{\rm x}=0.84\log\biggl[\nu F_{\nu}(1528\ {\rm \AA})/\nu F_{\nu}(2\ {\rm keV})\biggr]+0.85\biggr)$.
The {\it solid} line is the best fit line to 
Figure 8 of \citet{walt93}, and the {\it dashed} line is a chi-by-eye
fit to the outliers in the WF93 sample.  The {\it dotted} line
divides the ``normal'' sample of AGN from the outliers listed
in Tab. \ref{tabOutliers}.  Our data indicate either weaker 
UV emission or steeper $\alpha_{\rm x}$ among our sample compared
to WF93.
The left panel compares the cumulative distributions of $\alpha_{\rm x}$
for the NLS1 ({\it dashed}) and BLS1 ({\it solid}) samples.
\label{figSlope}}
\end{figure}

\begin{figure}
\epsscale{1.0}
\plotone{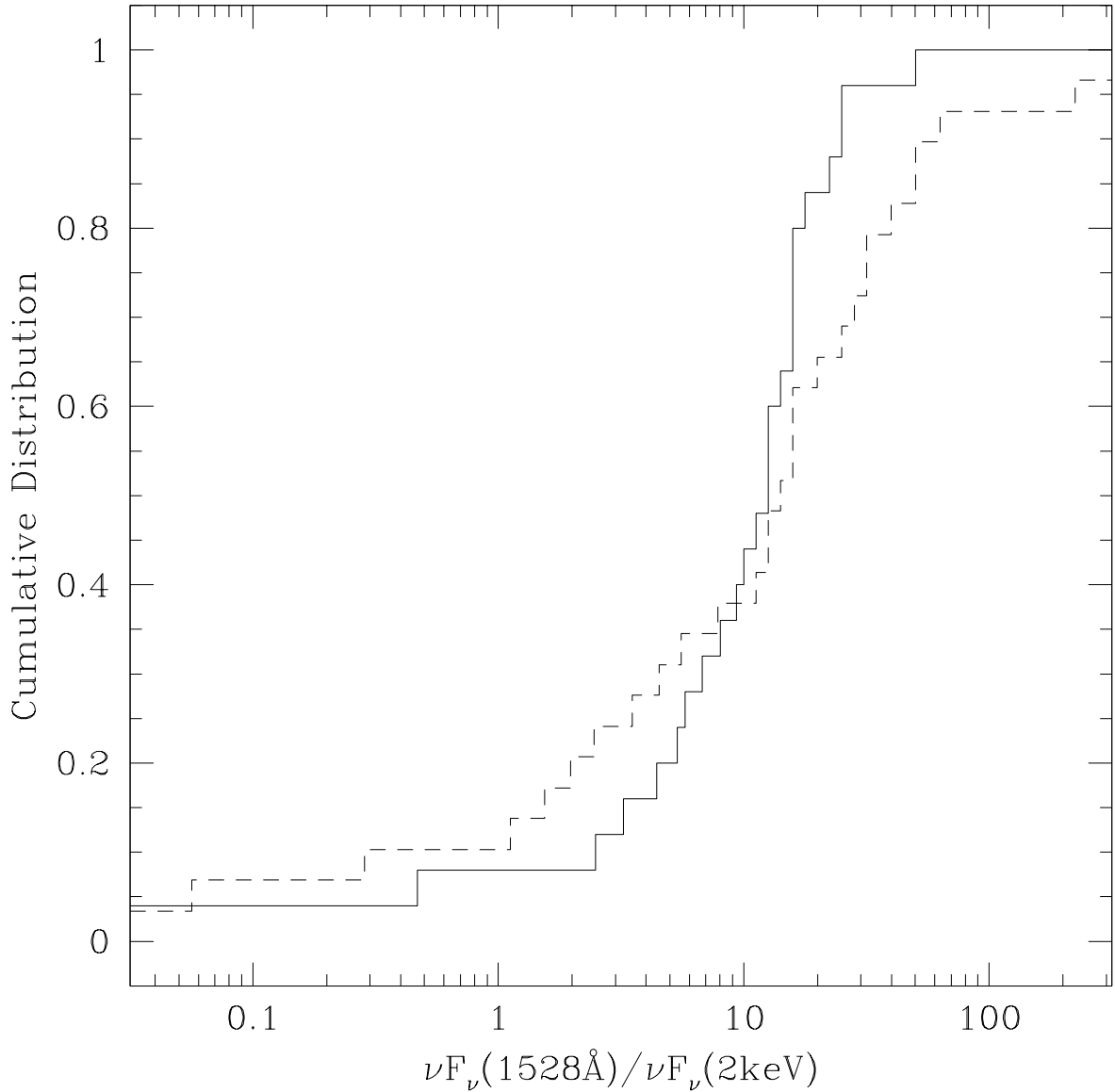}
\caption{Comparison of the distributions of BBB strength for the BLS1 
({\it solid}) and NLS1 ({\it dashed}) samples.  The two distributions are
obviously similar near the median values (CDF=0.5), but differ significantly
in the wings.  The KS test using the Kuiper significance criterion yields
a probability of 1.5\% that the NLS1 and BLS1 samples are drawn from
the same underlying population.
\label{figBumpCumulative}}
\end{figure}

In Figure \ref{figSlope}, we compare an indicator for the 
strength of the BBB with respect to the hard X-ray continuum 
$\biggl(\nu F_{\nu}(1528\ {\rm \AA})/\nu F_{\nu}(2\ {\rm keV}) \biggr)$ with
$\alpha_{\rm x}$ to verify that the correlation between the strength
of the BBB and the soft excess, as reported by WF93, also appears in
our sample.
We find a significant correlation in both the NLS1 and
BLS1 samples as well as in the merged
sample, as indicated in Table \ref{tabFluxCorrelation}.  
It is apparent that the majority of the BLS1s lie on or near the
WF93 relation, but the NLS1s are located
systematically above the WF93 best-fit power law.
Computing the best-fit relation to our data points in the figure yields
\begin{equation}\label{eqAlphaFit}
\alpha_{\rm x}=(0.84\pm0.13)\log\biggl[\nu F_{\nu}(1528\ {\rm \AA})/\nu F_{\nu}(2\ {\rm keV})\biggr]+(0.85\pm0.15)
\end{equation}
which is steeper than the WF93 best-fit, which has slope $0.68\pm0.1$.  The
two fits overlap in the regime occupied by the BLS1s, and the steeper slope
of our fit is driven by the NLS1s in our sample.
A two-dimensional KS test confirms that the NLS1 and BLS1 samples 
occupy a different region in the parameter
space at the 99.5\% confidence.  
Figure \ref{figBumpCumulative} shows
that the NLS1 and BLS1 samples occupy similar ranges in bump strength,
but the NLS1 sample shows extended tails at both ends.  

\begin{figure}
\epsscale{1.0}
\plotone{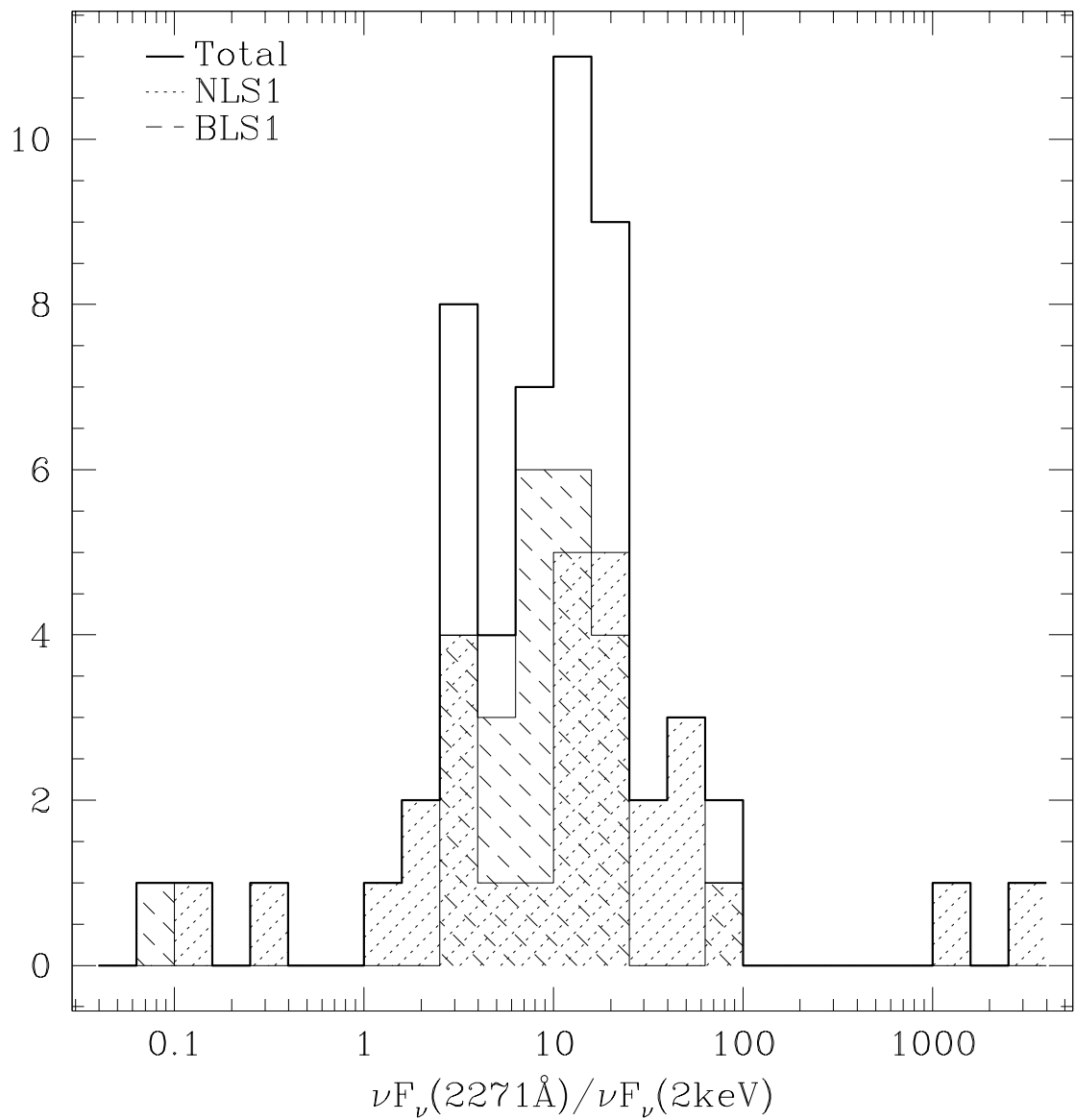}
\caption{Distributions of $NUV$ to X-ray flux ratios among the NLS1
({\it dotted fill}) and BLS1 ({\it dashed fill}) samples.  A flux ratio of 10
corresponds to $\alpha_{\rm ox}\approx1.4$.  A KS-test yields a 5\%
probability that the two samples are drawn from the same parent population.
\label{figAlphaOx}}
\end{figure}

We also computed the ratio 
$\biggl(\nu F_{\nu}(2271\ {\rm \AA})/\nu F_{\nu}(2\ {\rm keV})\biggr)$ between
the $NUV$ and X-ray fluxes, which is analogous to $\alpha_{\rm ox}$.
The distributions of the flux ratio in both the NLS1 and BLS1 
samples are shown in Figure \ref{figAlphaOx}.  It is apparent
that the NLS1 sample has more objects with high flux ratios,
consistent with Figure \ref{figBumpCumulative}, and a KS test
indicates that the two distributions are different at about 97\%
confidence, which is suggestive but not especially significant.
If we assume the UV continuum is well-described by a power-law, we
can determine $\alpha_{\rm ox}$ for our AGNs by comparing the $FUV$
and $NUV$ fluxes.  We find that
the mean and median of the $\alpha_{\rm ox}$ distribution of the
sample are both 1.4, consistent with the results of \citet{elvi94}.
However, we caution that this calculation requires extrapolating the
UV power-law longward of the $NUV$ effective wavelength, rendering
$\alpha_{\rm ox}$ inherently less robust than the
flux ratios shown in Figure \ref{figAlphaOx}.
In either case, our AGNs appear to be quite
typical in this respect, but there is marginal evidence that the
NLS1s have slightly stronger BBB than usual, consistent with the
results in Figure \ref{figSlope}.  If BBB photons are
reprocessed to form the soft excess, a stronger BBB should be associated
with a stronger soft excess, which indeed is observed.  (In order to
explain the flux ratios in Figure \ref{figSlope}, only one in
$\sim10^{6}$ BBB photons needs to be reprocessed.  The associated UV flux
decrement would not be observable.)

\begin{figure}
\epsscale{1.0}
\plotone{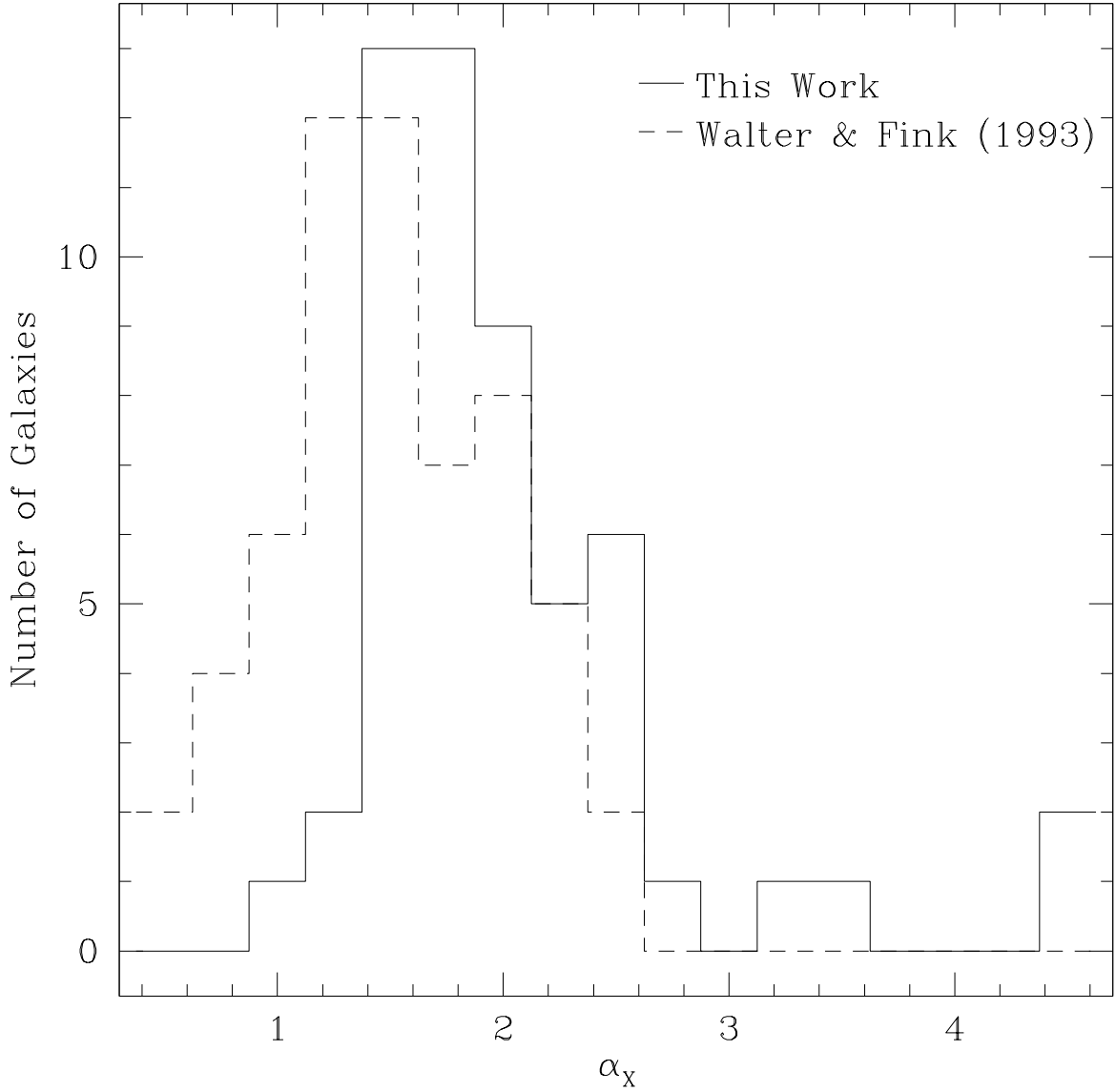}
\caption{Comparison between X-ray spectral indices in \citet{walt93} with the
galaxies we examine.  It is apparent that there is a significant preference
for steeper soft X-ray spectra in our sample compared to that of
WF93.  Our sample also contains more objects with
unusually soft X-ray spectra.
\label{figSlopeCompare}}
\end{figure}

Many of our AGNs occupy the gap between the main locus of 
WF93 galaxies and their outliers, as shown in Figure
\ref{figSlope}, indicating a systematic
difference between our AGNs and those of WF93.
This difference could be caused either by weaker UV at fixed
X-ray flux or by steeper $\alpha_{\rm x}$ at fixed BBB strength.  
It is apparent from
Figure \ref{figSlopeCompare} that the X-ray spectra of our galaxies are, on
average, steeper than the galaxies of WF93 
($\langle\alpha_{\rm x}\rangle=2.1$, compared to 
$\langle\alpha_{\rm x}\rangle=1.5$
for WF93).  Comparing the far-UV and X-ray fluxes of our sample with
the fluxes of the WF93 AGNs, we find that our AGNs are fainter in both
the UV and X-ray, but the difference is larger in the UV. (The median
is shifted by a factor of 5.7 in the UV, compared to 5.0 in the X-ray.)
This, in combination with the higher average $\alpha_{\rm x}$ in our
sample accounts for the observed differences between our sample and
WF93's.  The difference between the median 
$\nu F_{\nu}(1528\ {\rm \AA})/\nu F_{\nu}(2\ {\rm keV})$ in our sample and
WF93's might be attributable to their use of IUE observations to measure
UV fluxes.  The WF93 fluxes show a fractional error near unity for fluxes below
$\sim3\times10^{-11}$, whereas the typical GALEX uncertainty is only a few
percent at these flux levels.  We are therefore able to obtain
significant GALEX detections of all of our sources, and our sample
is unbiased with respect to UV flux.

While the majority of the BLS1s agree well with the WF93 best fit, the
NLS1s are shifted systematically to higher $\alpha_{\rm x}$.
In combination with the good correlation between
$\nu F_{\nu}(1528\ {\rm \AA})/\nu F_{\nu}(2\ {\rm keV})$ and 
$\alpha_{\rm x}$ for the full
sample, this indicates that the relation between the strength of the
BBB and the shape of the soft X-ray continuum is steeper among AGN with
the strongest soft excesses.  This in turn suggests the need
for a second parameter to account for the variation in $\alpha_{\rm x}$
at fixed BBB strength.

We identified 9 AGN that lie well away from the ``main''
relation between $\nu F_{\nu}(1528\ {\rm \AA})/\nu F_{\nu}(2\ {\rm keV})$
and $\alpha_{\rm x}$.  These outliers
are listed, along with a number of important properties, in Table 
\ref{tabOutliers}.  Five of the outliers show UV-optical luminosity
ratios less than one, putting them well below the main locus in
Figure \ref{figLuminosity}.  This suggests that the primary cause
of our outliers is UV absorption.  We fit a power law to the objects
in Figure \ref{figLuminosity} with $L_{FUV}\geq L_{V}$ and found,
\begin{equation}
\log(L_{FUV}/L_{\odot})=1.09\log(L_{V}/L_{\odot})-0.65
\end{equation}
Assuming that all of the galaxies with $L_{FUV}<L_{V}$ fall exactly
on the best fit line, we require internal E(B-V) between 0.3 and 1.0
to explain the measured luminosity ratios.  For the Galactic gas-to-dust
ratio, this implies 
$N_{H}\approx5\times10^{21}\ {\rm cm^{-2}}$, which is far larger than the
column densities measured from X-ray spectral fits.
We note, however, that the implied ${\rm N_{H}}$
is degenerate with $\alpha_{\rm x}$, so these systems could have larger
column densities and steeper spectra than reported, though this seems
unlikely given the recorded values of $\alpha_{\rm x}$.
Also, WF93 found a small number of galaxies with significant
internal extinction despite moderate column densities inferred
from the ROSAT spectra of those systems.
The unusual luminosity ratios shown in Figure \ref{figLuminosity} 
might also be an indication that the
$FUV$ and $V$ band luminosities are dominated by young stars rather than
by the AGN accretion disk.  This hypothesis is supported by the unusually
low values of $\alpha_{\rm ox}$ exhibited by three of these five objects.
All three AGNs with such low $\alpha_{\rm ox}$ have 
$L_{FUV}<10^{43} {\rm erg\ s^{-1}}$, which corresponds to a SFR of 
$10\ M_{\odot}\ {\rm yr^{-1}}$ \citep{sali07}.

Of the 4 outliers in Table \ref{tabOutliers}
that do not appear to be strongly absorbed in Figure \ref{figLuminosity},
3 show unusually large $\alpha_{\rm x}$, and the fourth
(RX J0902-07) lies very
close to the line dividing the ``normal'' AGNs from the outliers.  
This last object does not differ significantly from the ``typical'' 
AGNs in our sample for any of the parameters listed in Table \ref{tabOutliers},
suggesting that it should be considered normal.
The other outliers can be divided into two classes:
objects that show UV absorption and objects
that show extraordinarily high $\alpha_{\rm x}$.

Most of our AGNs 
occupy the gap between the WF93 best-fit relation
and their outliers.  \citet{walt93} explained
their outliers as normal objects with strong intrinsic absorption,
but few of our AGNs show evidence for UV absorption.  Only 3 of the 8 objects
with $L_{FUV}/L_{V}<1$ fall into our main sample, so strong UV absorption
cannot be responsible for this difference between our sample and WF93's.
We note, however, that weak absorption is difficult to
identify from Figure \ref{figLuminosity} due to the large intrinsic
scatter about the mean relation, so weak intrinsic absorption might contribute
to the shift in our sample away from the WF93 mean.

We also examine the relation between indicators for the strength 
of the BBB and its shape 
$\biggl(\nu F_{\nu}(1528\ {\rm \AA})/\nu F_{\nu}(2271\ {\rm \AA})\biggr)$,
shown in Figure \ref{figXrayShape}.  Like WF93, we find a
plateau accompanied by a sharp drop toward lower values
of $\nu F_{\nu}(1528\ {\rm \AA})/\nu F_{\nu}(2\ {\rm keV})$. However,
Figure \ref{figXrayShape}
shows a broader scatter in the plateau region, plateaus at a lower
ratio, and fills in the red tail of the distribution 
less completely than seen in the analogous diagram in WF93.  Also,
the objects occupying the ``tail'' of the
distribution in Figure \ref{figXrayShape} are all listed as outliers
in Table \ref{tabOutliers}, which immediately suggests that
the tail in our sample is due to absorption.  
The correlation between the strength and shape of the BBB
disappears if we disregard the outliers, indicating that the 
shape of the BBB is largely independent of its strength.

The absence of any correlation between the shape and strength of the
BBB is a direct contradiction of the results of WF93.  
We use very different methods to measure
the shape parameter of the BBB, so it is possible that the disparity
between our results and theirs are due to systematic biases, particularly
since our wavelength baseline is only half theirs.
However, the average UV flux ratio in Figure \ref{figXrayShape} 
$\biggl(\langle\nu F_{\nu}(1528\ {\rm \AA})/\nu F_{\nu}(2271\ {\rm \AA})\rangle\approx1.4\biggr)$
implies a power-law continuum ($F_{\nu}\propto\nu^{-\alpha_{uv}}$)
with $\langle\alpha_{uv}\rangle\approx-0.85$, which in
turn suggests that 
$\langle\nu F_{\nu}(1375\ {\rm \AA})/\nu F_{\nu}(2675\ {\rm \AA})\rangle\approx1.75$.
This is consistent with the plateau seen in WF93 Figure
11, which is at approximately 1.8.  
Given this agreement and the fact that the tail
of our distribution is populated by AGNs showing probable absorption, we
suggest that WF93 were too quick to dismiss absorption as a
potential cause of their correlation.

To verify that the differences between Figure \ref{figXrayShape} and
WF93's Figure 11 are not caused by systematic differences
between $\nu F_{\nu}(1528{\rm \AA})/\nu F_{\nu}(2\ {\rm keV})$ and
$\nu F_{\nu}(1528\AA)/\nu F_{\nu}(25\mu m)$,
we compare the two strength indicators in Figure \ref{figBumpStrength}.
Despite the differences between the UV fluxes of the two samples,
our galaxies show good agreement with the WF93 best-fit relation.
We derive a best fit relation for our sample, obtaining
\begin{equation}
\frac{\nu F_{\nu}(1528\ {\rm \AA})}{\nu F_{\nu}(25\ {\rm \mu m})}=(0.95\pm0.08)\frac{\nu F_{\nu}(1528\ {\rm \AA})}{\nu F_{\nu}(2\ {\rm keV})}-(1.16\pm0.09)
\end{equation}
which is consistent with the WF93 best fit within the uncertainties.
Thus, there is no inherent bias in 
$\nu F_{\nu}(1528\ {\rm \AA})/\nu F_{\nu}(2\ {\rm keV})$ compared to
$\nu F_{\nu}(1528\ {\rm \AA})/\nu F_{\nu}(25\ {\rm \mu m})$,
and the absence of a correlation
between the shape and strength of the BBB among our AGN sample
is not caused by differences between our strength indicator
and WF93's.  It also indicates that the
steeper relation between $\alpha_{\rm x}$ and
$\nu F_{\nu}(1528\ {\rm \AA})/\nu F_{\nu}(2\ {\rm keV})$ among our AGN compared
to the WF93 sample is not due to systematic errors.
This lends credence to the hypothesis that a factor besides the
strength of the BBB must
contribute to the shape of the soft X-ray continuum.

The structure of a standard thin disk, which might be reflected
in the UV color of the disk, can depend on both black
hole mass and Eddington ratio (see Eq. \ref{eqTemp}), so
We want to know whether
$\nu F_{\nu}(1528\ {\rm \AA})/\nu F_{\nu}(2271\ {\rm \AA})$ shows
systematic differences between the NLS1 and
BLS1 samples.  A KS test reveals that the distributions
differ between the NLS1
and BLS1 samples at about 95\% confidence, but this difference disappears
when we eliminate the
outliers (Tab. \ref{tabOutliers}).  Thus,
we measure no intrinsic variation in the structure of accretion disks
powering NLS1 and BLS1 AGNs.
However, the GALEX bands are sensitive only to variations in disk structure
if the Eddington ratio is well below $\dot{m}$.  (See Figure 
\ref{figShapeMass}.)  As a result, we expect little intrinsic difference
between the BLS1 and NLS1 AGN samples.

Finally, we look for any relationships of the indicators we have already
examined with FWHM(H$\beta$), which might tend to indicate systematic 
differences between the two classes of AGN.
We find two trends that might be of interest: a correlation with
$\alpha_{\rm x}$ among the merged AGN sample at
$>99.9$\% confidence and another with
$\nu F_{\nu}(1528\ {\rm \AA})/\nu F_{\nu}(2271\ {\rm \AA})$
among the NLS1 sample at 97\%
confidence.  The second is interesting if true, because it would suggest
that the structure of the BLR is related to the UV color of the accretion
disk, but the correlation is not strong enough to
support such a claim unequivocally.  Figure \ref{figWidthCorr}
shows the relation between FWHM(H$\beta$) and $\alpha_{\rm x}$ and
is consistent with the ``zone of avoidance,'' in which BLS1s generally have
$\alpha_{\rm x}\lesssim2.0$, as reported by \citet{boll96}.
The measured correlation is a result of this effect.

\subsection{Physical Parameters}\label{secPhysical}
We also examined the relationships between the observables discussed
above and the physical parameters ($M_{BH}$, $\dot{m}=L/L_{\rm edd}$
and $L_{\rm bol}$) that characterize each AGN, where the Eddington ratios were
all determined using $L_{\rm bol}$ from the Grupe et al. catalogs.
We find strong correlations ($>99$\% confidence) 
of shape, strength and $\alpha_{\rm x}$ with UV luminosity among the NLS1 
sample, as shown in Figure \ref{figLuminosityCorr}.  We also find
correlations of both strength and shape with luminosity 
in the merged sample, but the lack of correlation of $\alpha_{\rm x}$ 
with luminosity among BLS1s
dilutes the correlation in the merged sample to 98\% confidence.
This is significantly weaker than the same correlation
reported by \citet{kell08} for a sample of 
optically-selected, radio-quiet quasars.  This suggests that the 
shape of the X-ray continuum is more tightly coupled to the Big Blue
Bump among optically selected AGN than among X-ray selected AGN.  The
Spearman correlation coefficients and significance values for the various
parameters we examined are listed in Table \ref{tabPhysicalCorrelation}.

The strong positive correlation of 
$\nu F_{\nu}(1528\ {\rm \AA})/\nu F_{\nu}(2271\ {\rm \AA})$ with
$\nu L_{\nu}(1528\ {\rm \AA})$ is in conflict with the 
results of \citet{scot04}, who reported
that the BBB, as measured in the FUSE band (900-1200 \AA)
becomes softer in more luminous AGN.  However, the GALEX $FUV$ band
does not overlap with the FUSE coverage, so the different variations
with luminosity might be strictly a wavelength effect.  If this is
the case, the peak of the BBB in the average Seyfert
AGN must lie somewhere between 1500\AA and 900\AA.
The observed correlations
are weaker but still significant ($>99$\% confidence) if we consider
$\nu L_{\nu}(5100\ {\rm \AA})$.  In Figures \ref{figLuminosityCorr}a and 
\ref{figLuminosityCorr}b, the relations exhibited by the NLS1 and BLS1
show good agreement, which is consistent with the structure of the accretion
disks showing little variation between the two classes.
We see very different relations of $\alpha_{\rm x}$ with
luminosity between the NLS1 and BLS1 samples.  This could be caused by
most NLS1s being within a factor
of a few in $M_{BH}$, causing variations in $\dot{m}$ to drive
a correlation of $\alpha_{\rm x}$ with luminosity.

From simple virial considerations, we expect that the width
of the H$\beta$ emission line should correlate with both $M_{BH}$
and $\dot{m}$.  If we assume the simplest possible
relation, $R_{BLR}\propto L^{1/2}$, which is consistent with the
results of \citet{bent06}, we find that 
${\rm w(H}\beta{\rm )}\propto (M_{BH}/\dot{m})^{1/4}$.
Examining the correlations in Table \ref{tabPhysicalCorrelation}, we find that
the H$\beta$ line width is indeed correlated with both $M_{BH}$
and $\dot{m}$, but the correlation is much stronger with
$M_{BH}/\dot{m}$.  Fitting the FWHM to $M_{BH}/\dot{m}$ yields
\begin{equation}
\log\biggl[{\rm w(H}\beta)\biggr]=(0.24\pm0.01)\log(M_{BH}/\dot{m})+(1.5\pm0.1)
\end{equation}
with $\chi^{2}_{\nu}=1.02$,
which is consistent with the simple prediction above.
This relationship is also consistent with the
results of \citet{mcha06}, who found that the break timescale of the
power density spectrum, which is proportional to
$M_{BH}/\dot{m}$, is well correlated with line width.  This good agreement
both with theory and with previous observations suggests that the
\citet{grup04c} mass measurements are, on average, robust.

Our correlation measurements also agree with \citet{pico05}, who found a strong 
anti-correlation of $\Gamma_{\rm soft}$ with H$\beta$ line width
($r_{s}=-0.54$),
where $N_{\nu}\propto\nu^{-\Gamma_{\rm soft}}$ is measured
from 0.3--2.0 keV.
In fact, the strength of the correlation they report
is very similar to the strength of the correlation we find between 
FWHM and $\alpha_{\rm x}$ (our $r_{s}=-0.53$),
suggesting that much of the scatter between the two variables
may be intrinsic.  The errors on the line widths of NLS1s are comparable
to the errors on BLS1 widths, so the lack of correlations with
$M_{BH}$ or $\dot{m}$ among the BLS1 sample suggests
that the geometry factor $f$ varies more among the BLS1 sample than the 
NLS1 sample.  This might happen if the BLR becomes
more spherically symmetric at high Eddington ratio.

There is also a correlation of moderate significance (98.4\% confidence)
between $M_{BH}$ and
$\nu F_{\nu}(1528\ {\rm \AA})/\nu F_{\nu}(2271\ {\rm \AA})$, but correlation
is positive, which is in opposition to the trend predicted
for a standard thin disk (see Fig. \ref{figShapeMass}).
This correlation
is probably spurious, since it is driven by
a small number of AGNs with 
$\nu F_{\nu}(1528\ {\rm \AA})/\nu F_{\nu}(2271\ {\rm \AA})<1$, three of which
are outliers in Figure \ref{figSlope}.  Two additional AGNs with
$\nu F_{\nu}(1528\ {\rm \AA})/\nu F_{\nu}(2271\ {\rm \AA})<1$ have
$L_{FUV}<\lambda L_{\lambda}(5100\ {\rm \AA})$, suggesting
that low levels of
recent star formation could significantly influence the measured
UV flux ratio.  After excluding
both of these groups, we find no significant correlation of
$\nu F_{\nu}(1528\ {\rm \AA})/\nu F_{\nu}(2271\ {\rm \AA})$ with $M_{BH}$.
There is no correlation of 
$\nu F_{\nu}(1528\ {\rm \AA})/\nu F_{\nu}(2271\ {\rm \AA})$
with $\dot{m}$ regardless of whether the AGNs with
$\nu F_{\nu}(1528\ {\rm \AA})/\nu F_{\nu}(2271\ {\rm \AA})<1$ are considered,
which is expected given the relatively high Eddington ratios
typical of the AGNs in our sample.

\section{Theoretical Models}\label{secModels}
We would like to use the UV properties of the observed AGNs to
place constraints on theoretical models for the soft X-ray excess.  We
therefore compare the UV properties of our sample to
predictions from various models.
For the standard Shakura-Sunyaev thin disk,
the temperature at the inner edge of the disk 
is given by their Eq. 3.8,
\begin{equation}\label{eqTemp}
T_{\rm inner}=5\times10^{8}{\rm K\ } \alpha^{1/5} \dot{m}^{4/5} m^{-1/5} r_{ls}^{-3/2} \left(1-r_{ls}^{-1/2}\right)^{4/5}
\end{equation}
where $m=M/M_{\odot}$ and 
$r_{ls}$ is the last stable radius in units of the Schwarzschild
radius.  We computed $T_{\rm inner}$ for all the AGNs
displayed in Figure \ref{figShapeMass}, assuming $\alpha=0.1$, 
$r_{ls}=1.5$ and $\dot{m}$ determined by the measured luminosities and
black hole masses.  We found that only NGC 7214 has
a maximum disk temperature near $(1+z)T_{FUV}$.  We note
that NGC 7214 falls in the ``main relation'' in Figure \ref{figShapeMass}
and has $\nu F_{\nu}(1528\ {\rm \AA})/\nu F_{\nu}(2271\ {\rm \AA})>1$, 
meaning it is
essentially normal.  This is consistent with our argument that GALEX is
insensitive to changes in the structure of the accretion disk.
The accretion rates implied by the UV colors of our AGNs 
(Fig. \ref{figShapeMass}) are substantially
lower than the accretion rates determined using the bolometric luminosities
from the literature ($\dot{m}\gtrsim0.1$).  This
implies that either the measured UV fluxes suffer from substantial intrinsic
extinction, which would naturally redden the emergent spectrum,
or the $NUV$ fluxes might contain a substantial
contribution from sources other than the accretion disk.  \citet{walt93}
are able to measure the Balmer decrements for their sources, and they find
that most of their sample suffers from little to no intrinsic extinction.
Given the excellent agreement between the UV and MIR properties of their
sample and ours (Figure \ref{figBumpStrength}), intrinsic
extinction is highly unlikely to influence our results.  The most likely source
of a significant contribution to the $NUV$ fluxes of our objects is
low-level star formation, but a non-standard disk
structure is also possible.

A popular class of models for non-standard accretion disks is the slim disk
model first proposed by \citet{much82}, in which super-Eddington accretion
drives the disk to puff up and change its structure.
The super-Eddington accretion rates observed in several
of our AGN suggest that this model might be applicable.
Even more objects move into the slim disk regime if
we consider the ``corrected'' black hole masses rather than the virial
products, because the geometric factor for masses determined
using the FWHM of the $H\beta$ emission line is less than 1 
(e.g. \citealt{wats07}).  The slim disk models of \citet{wang03} predict
that the SED of the BBB rises steeply toward higher energy in the UV
for even moderately super-Eddington accretion ($\dot{m}\gtrsim5$), so
$\nu F_{\nu}(1528\ {\rm \AA})/\nu F_{\nu}(2271\ {\rm \AA})$ 
should be greater than 1.
This is generally true for our sample, but it is unable to 
explain the most unusual objects in Figure \ref{figShapeMass},
which have UV flux ratios lying below rather than above the predictions
of a standard thin disk.
Furthermore, the slim disk model predicts that $\alpha_{\rm x}$ 
should decrease slightly
with increasing $\dot{m}$, which has already been demonstrated to be false
(e.g. \citealt{grup04b}).  Alternative slim disk models from \citet{kawa03} 
and \citet{chen04} show similar failings.
Furthermore, two of the most extreme objects in Figure \ref{figShapeMass} are
BLS1s and show moderate Eddington ratios ($\dot{m}=0.08, 0.06$ respectively).
This precludes the application of slim disk theory to these objects.

Models that rely on Comptonization of thermal
photons in a hot plasma (e.g. \citealt{kawa01}) are motivated 
by the strong correlation between 
$\nu F_{\nu}(1375\ {\rm \AA})/\nu F_{\nu}(2\ {\rm keV})$ 
and $\alpha_{\rm x}$ reported by WF93.  
Because our sample differs systematically from the WF93 best-fit,
we infer that an additional parameter
related to the strength of the soft excess may influence the
$\nu F_{\nu}(1528\ {\rm \AA})/\nu F_{\nu}(2\ {\rm keV})$---$\alpha_{\rm x}$ 
relation.  However, the underlying model is supported by the strong
correlation between the strength of the BBB and the shape of the
soft X-ray continuum in our data.
Given the significantly different
$\langle M_{BH}\rangle$ and $\langle \dot{m}\rangle$
exhibited by BLS1s and NLS1s, it is
logical to infer that the temperature or density of the disk corona
might be responsible for the different 
$\nu F_{\nu}(1528\ {\rm \AA})/\nu F_{\nu}(2\ {\rm keV})$---$\alpha_{\rm x}$
relations exhibited by the two samples, since both $M_{BH}$ and $\dot{m}$
can influence the structure of the corona.  Since the BLS1s, on average,
agree well with the WF93 results, this could also explain the differences
between our merged sample and WF93's results.

Models that suggest an atomic origin for the soft excess,
originally proposed by \citet{gier04}, postulate that the soft excess
is actually a ``hard deficit,'' in which the X-ray flux in the range
$0.7\ {\rm keV} \lesssim E \lesssim 5\ {\rm keV}$ is subject to 
significant absorption
by relativistically broadened O{\sc vii} and O{\sc viii} lines.
In a recent paper, \citet{schu07} modeled the emergent X-ray spectrum
that would be observed following
absorption by material in a UV line-driven wind.  They exclude
this model based on sharp absorption features that appear in the model
spectra but not in the spectra of real AGN.  They also suggest that this
could be resolved by invoking magnetically-driven outflow, which can
potentially reach much larger terminal velocities.

\citet{schu06} modeled the X-ray spectrum of
PG 1211+143 with absorption in a high-velocity wind without subjecting the
wind to any physical constraints.  We want to determine whether our
UV flux measurements are consistent with a smeared absorption model,
assuming a mechanism to drive an outflow with the necessary velocity
profile could be found.  Since the X-ray continua in soft
excess AGNs are generally smooth \citep{schu07}, we assume that the 
multiplicative flux
decrement from an input power-law continuum to the measured flux at
2 keV is linearly proportional to the strength of the soft X-ray excess.
To relate the strength of the soft excess to the soft X-ray spectral
index ($\alpha_{\rm x}$), we fit the soft excess and
$\Gamma_{\rm x}$ measurements of by WF93, finding
\begin{equation}\label{eqWFGamma}
\log(\Gamma_{\rm x})=(1.5\pm0.2)\log(X)+(1.9\pm0.1)
\end{equation}
with $\chi^{2}_{\nu}=0.59$,
where $\Gamma_{\rm x}$ is the photon spectral index in the ROSAT band, and
$X$ is the strength of the soft excess relative to the hard X-ray continuum,
following WF93.  Inverting this equation and transforming to $\alpha_{\rm x}$
yields:
\begin{equation}\label{eqStrength}
\log(X)=\frac{\alpha_{\rm x}-0.9}{1.5}
\end{equation}
We predict the flux decrements
required to produce the measured $\alpha_{\rm x}$ in each of our sources
using Equations \ref{eqWFGamma} and \ref{eqStrength} in combination with the 
rest-frame 2 keV flux decrement for PG 1211+143 required by Schurch \& Done
(2006; $\alpha_{\rm cont}=1.37$, flux decrement = 1.3 from their Figure 5).

To determine the minimum 
$\nu F_{\nu}(1528\ {\rm \AA})/\nu F_{\nu}(2\ {\rm keV})$
required by the model, we need to know the shape of the input continuum
for each AGN in our sample.
We compute the rest-frame $F_{\nu}(5100\ {\rm \AA})$ for each of our
AGNs from the published $L_{\nu}(5100\ {\rm \AA})$ in Grupe et al. 
(1998, 2004), using measured $F_{\nu}(2271\ {\rm \AA})$ to estimate
K-corrections and assuming a power-law continuum.
We calculate the continuum shape ($\alpha_{\rm cont}$) from the 
measured 5100 \AA and 2 keV fluxes for each of our AGNs,
\begin{equation}\label{eqContinuum}
\alpha_{\rm cont}=-0.343\biggl(\log[F_{\nu}(2\ {\rm keV})]-\log[F_{\nu}(5100\ {\rm \AA})]\biggr)
\end{equation}
where the $F_{\nu}$ are rest-frame fluxes, and the X-ray flux has
been corrected for the appropriate flux decrement.  Using $\alpha_{\rm cont}$
and the flux decrements required by our {\it ad hoc} model, we predict lower
limits on $\nu F_{\nu}(1528\ {\rm \AA})/\nu F_{\nu}(2\ {\rm keV})$ for
each AGN.

We show the lower limits and measured flux ratios for
our objects, excluding the outliers, in Figure \ref{figRatioLimits}.
The blue triangles mark the six objects 
(H0439-27, Mrk 141, MCG+08-23-067, RX J1319+52, NGC 7214, RX J2349-31)
whose lower limits exceed the measured flux ratios.
The cumulative deficit distribution of these objects, 
normalized to the uncertainties in their flux ratios,
is shown in Figure \ref{figLowerCum}.  This distribution is
consistent with all of the objects having flux ratios intrinsically
equal to the lower limits but scattered low by 
the observational errors.  Thus, we cannot rule out an origin of the
soft excess in smeared absorption based on our GALEX measurements.

While the AGNs in our sample are inconsistent with slim disk models,
our measured flux ratios are consistent with either smeared
absorption or Comptonization in a hot corona.  We are therefore unable
to favor either of these competing models, though the differences between
our best-fit 
$\nu F_{\nu}(1528\ {\rm \AA})/\nu F_{\nu}(2\ {\rm keV})$---$\alpha_{\rm x}$
relation and WF93's
indicates the need for a second parameter in the Comptonization model.  We
suggest this parameter might be the Eddington ratio.

\section{Summary \& Conclusions}\label{secConclusions}
We measure the UV fluxes of a sample of X-ray selected AGNs with
strong soft X-ray excesses.  We find that our AGNs are slightly fainter
in the UV compared to the X-ray than a similar sample studied by WF93, and
we conclude that these differences are attributable to selection effects.

We examine the relationships between several observables and the
inferred physical properties of our AGN.  We find that the shape
of the soft X-ray continuum shows significant
correlations with $\nu F_{\nu}(1528\ {\rm \AA})/\nu F_{\nu}(2\ {\rm keV})$,
but the slope of the relation is steeper than that measured by WF93.
This difference appears to result from selection effects.  
We conclude that the X-ray spectra of AGN with unusually steep
soft X-ray continua, which belong to the NLS1 class, 
are related to their UV spectra
in a way fundamentally similar to AGN with more mundane soft
X-ray spectra.  The mechanism that drives the 
$\nu F_{\nu}(1528\ {\rm \AA})/\nu F_{\nu}(2\ {\rm keV})$---$\alpha_{\rm x}$
correlation must also lead to steeper $\alpha_{\rm x}$ at fixed
$\nu F_{\nu}(1528\ {\rm \AA})/\nu F_{\nu}(2\ {\rm keV})$ among NLS1s.
The Eddington ratio might make a good choice for this second parameter,
since the differences between our results and WF93's are largest for
the NLS1 sample.

We find a positive correlation of moderate significance between
$M_{BH}$ with the shape of the UV continuum, but this correlation
disappears if we disregard objects lying far from in main locus in
$\nu F_{\nu}(1528\ {\rm \AA})/\nu F_{\nu}(2\ {\rm keV})$---$\alpha_{\rm x}$ 
space.  We also find no evidence for a correlation of
$\nu F_{\nu}(1528\ {\rm \AA})/\nu F_{\nu}(2271\ {\rm \AA})$ with 
$L/L_{\rm edd}$.  If the soft X-ray excess is caused by Comptonization
of BBB photons in the hot corona of the accretion disk,
a second parameter is needed to explain the large
intrinsic variation in $\alpha_{\rm x}$ at fixed BBB strength.
Because $\alpha_{\rm x}$ is known to depend strongly on $L/L_{\rm edd}$ 
while and the properties of the accretion disk vary only weakly
with accretion rate, $L/L_{\rm edd}$ is
the most obvious candidate.

We find no significant correlation between the color
and strength of the BBB, so either the luminosity of the
accretion disk relative to the underlying power law is independent 
of the temperature of the disk, or the
characteristic temperature of the typical Seyfert 1 galaxy is outside
the range where GALEX colors are sensitive 
($5{\rm eV}\lesssim kT\lesssim 10{\rm eV}$).
The latter hypothesis is more likely based on the limited predicted range
in GALEX colors for the black hole masses and accretion rates appropriate
for our sample.  Comparisons between predicted and measured flux ratios
also suggest that the GALEX fluxes include contamination from young stars,
obscuring any underlying correlation in objects with
$\nu L_{\nu}(1528\AA)\lesssim5\times10^{43}{\rm erg\ s^{-1}}$ 
(estimated ${\rm SFR}\lesssim10\ M_{\odot}\ {\rm yr^{-1}}$; \citealt{sali07}).
Among our sample, there are 7(3) of 54(45) AGN in this luminosity
range including (excluding)
the outliers, so the impact of UV emission from young stars on our main
conclusions will be small.  Resolving
the question of whether or not the shape and strength of the BBB are
independent will likely require UV spectroscopy from HST, which has the
resolution to separate host starlight from AGN emission and can
be used to correct the measured flux ratios for redshift.

Finally, we are unable to use the UV fluxes of the Grupe et al. AGNs to
distinguish between the absorption and Comptonization models for the
origin of the soft X-ray excess.  Resolving this question could have important
implications for our understanding of AGN feedback, but the {\it ad hoc}
model we use to estimate minimum 
$\nu F_{\nu}(1528\ {\rm \AA})/\nu F_{\nu}(2\ {\rm keV})$ ratios does
not provide sufficient predictive power to determine whether the absorption
model really agrees with the UV flux measurements.  Further study with
a more detailed model is needed.

\acknowledgements
We thank an anonymous referee for insightful and penetrating comments which
have significantly improved this paper.  We are also 
grateful to G.C. Dewangan for helpful comments
and to Dirk Grupe for illuminating discussion regarding
the ROSAT spectra used to construct his AGN samples.
We also wish to thank the GALEX collaboration and the
Space Telescope Science Institute for providing access to the UV
images used in this work.  GALEX is a NASA Small Explorer Class mission.

\clearpage
\begin{deluxetable}{cccccccccccc}
\tabletypesize{\scriptsize}
\rotate
\tablewidth{0pc}
\tablecaption{Seyfert 1 Galaxy Sample
\label{tabGalaxies}}
\tablehead{
\colhead{Name} & \colhead{$\alpha_{\rm 2000}$} & 
\colhead{$\delta_{\rm 2000}$} & \colhead{z} &
\colhead{${\rm N_{H}}$} & \colhead{$\alpha_{\rm x}$} & \colhead{E($B-V$)} & 
\colhead{FWHM(${\rm H}\beta$)} &
\colhead{$\nu F_{\nu}(1582\ {\rm \AA})]$} & 
\colhead{$\nu F_{\nu}(2271\ {\rm \AA})$} & \colhead{$F_{\nu}(2\ {\rm keV})$} &
\colhead{Exp. Time} \\
\colhead{} & \colhead{} & \colhead{} & \colhead{} &
\colhead{$[10^{20}{\rm cm^{-2}}]$} & \colhead{} & \colhead{} & 
\colhead{$[{\rm km\ s^{-1}}]$} &
\colhead{$[10^{-12}{\rm erg\ s^{-1}\ cm^{-2}}]$} &
\colhead{$[10^{-12}{\rm erg\ s^{-1}\ cm^{-2}}]$} &
\colhead{$[10^{-13}{\rm erg\ s^{-1}\ cm^{-2}\ keV^{-1}}]$} & \colhead{$[{\rm s}]$}}
\startdata
RX J0022-34 & 00:22:33.0 & -34:07:22 & 0.219 & 1.39 & $1.6\pm0.2$ & 0.013 & 
$4110\pm120$ & $9.52\pm0.40$ & $0.51\pm0.01$ & $3.6\pm2.0$ & 112 \\
QSO 0056-36 & 00:58:37.0 & -36:06:06 & 0.165 & 1.94 & $1.62\pm0.51$ & 0.014 & 
$4550\pm250$ & $27.3\pm0.7$ & $19.8\pm0.3$ & $5.44\pm2.47$ & 112 \\
 & & & & & & & & $27.2\pm0.8$ & $19.7\pm0.3$ & & 76 \\
RX J0100-51 & 01:00:27.0 & -51:13:55 & 0.062 & 2.42 & $1.75\pm0.52$ & 0.015 & 
$3190\pm630$ & $34.7\pm0.7$ & $27.4\pm0.3$ & $10.2\pm4.8$ & 118 \\
MS 0117-28 & 01:19:36.0 & -28:21:31 & 0.349 & 1.65 & $2.51\pm0.66$ & 0.017 & 
$1681\pm260$ & $15.4\pm0.5$ & $12.5\pm0.2$ & $1.51\pm1.45$ & 117 \\
RX J0134-42 & 01:34:17.0 & -42:58:27 & 0.237 & 1.59 & $6.7\pm2.6$ & 0.017 & 
$1160\pm80$ & $17.0\pm0.6$ & $16.7\pm0.3$ & $(3\times10^{-5})\pm10^{-2}$ & 82 \\
RX J0136-35 & 01:36:54.0 & -35:09:53 & 0.289 & 5.60 & $4.9\pm0.5$ & 0.016 & 
$1320\pm120$ & $7.15\pm0.34$ & $5.02\pm0.13$ & $(2.04\pm4.38)\times10^{-2}$ & 119 \\
RX J0148-27 & 01:48:22.0 & -27:58:26 & 0.121 & 1.50 & $2.62\pm0.30$ & 0.017 & 
$1030\pm100$ & $20.4\pm0.5$ & $17.2\pm0.2$ & $7.25\pm1.89$ & 148 \\
RX J0152-23 & 01:52:27.0 & -23:19:54 & 0.113 & 1.10 & $1.75\pm0.39$ & 0.012 & 
$2890\pm250$ & $18.2\pm0.6$ & $13.9\pm0.2$ & $6.20\pm1.68$ & 108 \\
Mrk 1048 & 02:34:37.8 & -08:47:16 & 0.042 & 2.90 & $1.67\pm0.43$ & 0.033 & 
$5670\pm160$ & $44.1\pm0.1$ & $35.08\pm0.06$ & $18.9\pm7.5$ & 3397 \\
 & & & & & & & & $51.0\pm0.2$ & $40.16\pm0.09$ & & 1550 \\
RX J0323-49 & 03:23:15.0 & -49:31:51 & 0.071 & 1.72 & $2.35\pm0.29$ & 0.017 & 
$1680\pm250$ & $1.24\pm0.15$ & $2.62\pm0.09$ & $6.04\pm1.54$ & 112 \\
Fairall 1116 & 03:51:42.0 & -40:28:00 & 0.059 & 3.84 & $1.87\pm0.53$ & 0.013 &
$4310\pm630$ & $28.9\pm0.7$ & $22.0\pm0.3$ & $8.78\pm4.32$ & 118 \\
RX J0435-46 & 04:35:14.0 & -46:15:33 & 0.070 & 1.80 & $2.2\pm0.3$ & 0.014 & 
$3820\pm240$ & $4.68\pm0.27$ & $4.57\pm0.13$ & $1.17\pm1.15$ & 112 \\
RX J0435-36 & 04:35:54.0 & -36:36:41 & 0.141 & 1.49 & $1.6\pm0.2$ & 0.013 & 
$6750\pm620$ & $7.44\pm0.27$ & $6.90\pm0.16$ & $3.63\pm1.27$ & 112 \\
H0439-27 & 04:41:22.5 & -27:08:20 & 0.084 & 2.50 & $1.30\pm1.57$ & 0.036 & 
$2550\pm150$ & $3.66\pm0.23$ & $5.40\pm0.13$ & $8.65\pm4.72$ & 108 \\
RX J0454-48 & 04:54:43.0 & -48:13:20 & 0.363 & 1.91 & $2.4\pm0.7$ & 0.011 &
$1970\pm200$ & $2.43\pm0.21$ & $2.40\pm0.09$ & $2.50\pm1.07$ & 117 \\
RX J0902-07 & 09:02:33.6 & -07:00:04 & 0.089 & 3.31 & $2.12\pm0.62$ & 0.037 & 
$1860\pm150$ & $3.33\pm0.25$ & $2.90\pm0.11$ & $4.05\pm2.06$ & 78 \\
RX J1005+43 & 10:05:42.0 & +43:32:41 & 0.178 & 1.08 & $2.15\pm0.46$ & 0.011 & 
$2990\pm120$ & $14.5\pm0.5$ & $12.3\pm0.2$ & $3.16\pm1.78$ & 116 \\
RX J1007+22 & 10:07:10.2 & +22:03:02 & 0.083 & 2.76 & $2.91\pm0.54$ & 0.031 & 
$2740\pm250$ & $2.84\pm0.21$ & $2.97\pm0.10$ & $7.28\pm3.42$ & 114 \\
CBS 126 & 10:13:03.0 & +35:51:24 & 0.079 & 1.41 & $1.62\pm0.33$ & 0.011 & 
$2980\pm200$ & $17.7\pm0.5$ & $14.8\pm0.2$ & $10.1\pm2.7$ & 110 \\
Mrk 141 & 10:19:13.0 & +63:58:03 & 0.042 & 1.07 & $1.84\pm0.58$ & 0.010 & 
$3600\pm110$ & $9.46\pm0.40$ & $10.2\pm0.2$ & $3.32\pm1.63$ & 119 \\
Mrk 142 & 10:25:31.0 & +51:40:35 & 0.045 & 1.18 & $2.10\pm0.27$ & 0.016 & 
$1620\pm120$ & $22.3\pm0.6$ & $16.7\pm0.2$ & $4.61\pm1.13$ & 117 \\
RX J1117+65 & 11:17:10.0 & +65:22:07 & 0.147 & 0.91 & $2.50\pm0.49$ & 0.012 & 
$1650\pm170$ & $5.71\pm0.25$ & $6.77\pm0.13$ & $1.91\pm1.17$ & 168 \\
PG 1115+407 & 11:18:30.4 & +40:25:55 & 0.154 & 1.91 & $1.81\pm0.66$ & 0.016 &
$1740\pm180$ & $24.7\pm0.7$ & $20.1\pm0.3$ & $2.48\pm1.46$ & 100 \\
Mrk 734 & 11:21:47.0 & +11:44:19 & 0.033 & 2.64 & $2.0\pm0.2$ & 0.032 & 
$2230\pm140$ & $47.1\pm0.7$ & $38.9\pm0.3$ & $2.49\pm1.16$ & 141 \\
MCG+08-23-067 & 12:36:51.2 & +45:39:05 & 0.030 & 1.37 & $1.36\pm0.53$ & 0.017 & 
$730\pm140$ & $3.91\pm0.27$ & $3.98\pm0.13$ & $6.45\pm3.21$ & 99 \\
IC 3599 & 12:37:41.0 & +26:42:28 & 0.021 & 3.77 & $3.37\pm0.21$ & 0.019 & 
$635\pm110$ & $(4.85\pm2.11)\times10^{-2}$ & $0.41\pm0.04$ &
$5.88\pm1.39$ & 100 \\
NGC 4593 & 12:39:39.4 & -05:20:39 & 0.009 & 2.33 & $1.19\pm0.42$ & 0.025 & 
$4910\pm300$ & $50.5\pm0.9$ & $55.8\pm0.4$ & $46.5\pm18.9$ & 117 \\
IRASF 1239+33 & 12:42:11.0 & +33:17:03 & 0.044 & 1.35 & $2.02\pm0.42$ & 0.019 & 
$1640\pm250$ & $1.14\pm0.13$ & $1.79\pm0.08$ & $2.87\pm0.86$ & 112 \\
PG 1244+026 & 12:46:35.2 & +02:22:09 & 0.049 & 1.75 & $1.44\pm0.53$ & 0.026 & 
$830\pm50$ & $12.9\pm0.1$ & $11.84\pm0.04$ & $8.39\pm4.41$ & 2173 \\
RX J1304+05 & 13:04:17.0 & +02:05:37 & 0.229 & 1.77 & $2.26\pm0.59$ & 0.024 & 
$1300\pm800$ & $5.12\pm0.27$ & $3.44\pm0.11$ & $1.72\pm1.05$ & 113 \\
RX J1312+26 & 13:12:59.0 & +26:28:27 & 0.061 & 1.10 & $1.5\pm0.2$ & 0.012 & 
$2905\pm220$ & $5.57\pm0.34$ & $4.47\pm0.14$ & $4.48\pm2.45$ & 91 \\
RX J1319+52 & 13:19:57.1 & +52:35:33 & 0.092 & 1.19 & $2.06\pm0.44$ & 0.016 & 
$950\pm100$ & $1.15\pm0.15$ & $1.27\pm0.07$ & $4.68\pm2.50$ & 108 \\
RX J1355+56 & 13:55:17.0 & +56:12:45 & 0.122 & 1.15 & $2.31\pm0.42$ & 0.008 & 
$1110\pm100$ & $6.62\pm0.38$ & $7.20\pm0.17$ & $2.55\pm1.52$ & 89 \\
RX J1413+70 & 14:13:37.0 & +70:29:51 & 0.107 & 1.93 & $1.07\pm0.47$ & 0.016 & 
$4400\pm1000$ & $0.21\pm0.06$ & $0.18\pm0.03$ & $10.3\pm2.8$ & 110 \\
Mrk 478 & 14:42:08.0 & +35:26:23 & 0.077 & 1.04 & $2.22\pm0.16$ & 0.014 & 
$1630\pm150$ & $45.3\pm1.0$ & $39.0\pm0.4$ & $7.04\pm1.76$ & 81 \\
SBS 1527+56 & 15:29:07.5 & +56:16:07 & 0.100 & 1.29 & 1.46\tablenotemark{a} & 0.011 & 
$2760\pm420$ & $23.7\pm0.8$ & $14.1\pm0.3$ & $7.37\pm3.61$ & 64 \\
KUG 1618+40 & 16:19:51.3 & +40:58:48 & 0.038 & 0.93 & $1.87\pm0.45$ & 0.007& 
$1820\pm100$ & $4.86\pm0.34$ & $3.28\pm0.13$ & $3.90\pm2.30$ & 83 \\
EXO 1627+40 & 16:29:01.3 & +40:08:00 & 0.272 & 0.85 & $2.15\pm0.37$ & 0.009& 
$1450\pm200$ & $4.33\pm0.32$ & $3.80\pm0.12$ & $1.25\pm0.77$ & 83 \\
RX J2144-39 & 21:44:49.2 & -39:49:01 & 0.140 & 4.89 & $3.4\pm0.3$ & 0.023 & 
$1445\pm120$ & $1.30\pm0.17$ & $1.45\pm0.08$ & $2.58\pm1.99$ & 81 \\
NGC 7214 & 22:09:07.6 & -27:48:36 & 0.023 & 1.64 & $1.02\pm0.63$ & 0.019& 
$4700\pm250$ & $20.7\pm0.5$ & $17.8\pm0.2$ & $10.4\pm5.2$ & 113 \\
RX J2213-17 & 22:13:00.0 & -17:10:18 & 0.146 & 2.48 & $2.4\pm0.4$ & 0.026 & 
$1625\pm200$ & $5.29\pm0.27$ & $4.44\pm0.12$ & $1.51\pm1.90$ & 115 \\
RX J2216-44 & 22:16:53.0 & -44:51:57 & 0.136 & 2.17 & $1.98\pm0.38$ & 0.018 & 
$1630\pm130$ & $19.1\pm0.5$ & $14.0\pm0.2$ & $3.13\pm1.72$ & 118 \\
PKS 2227-399 & 22:30:40.3 & -39:42:52 & 0.318 & 1.25 & $0.72\pm0.55$ & 0.018& 
$3710\pm1500$ & $7.66\pm0.38$ & $5.82\pm0.16$ & $10.6\pm5.1$ & 94 \\
RX J2232-41 & 22:32:43.0 & -41:34:37 & 0.075 & 1.60 & $1.8\pm0.5$ & 0.013 & 
$4490\pm350$ & $3.50\pm0.23$ & $2.74\pm0.10$ & $1.31\pm2.46$ & 116 \\
RX J2241-44 & 22:41:56.0 & -44:04:55 & 0.545 & 1.76 & $2.5\pm0.4$ & 0.011 & 
$1890\pm200$ & $11.7\pm0.4$ & $12.5\pm0.2$ & $0.68\pm1.00$ & 118 \\
RX J2242-38 & 22:42:38.0 & -38:45:17 & 0.221 & 1.18 & $2.92\pm0.70$ & 0.014 & 
$1900\pm200$ & $8.35\pm0.40$ & $5.46\pm0.15$ & $1.22\pm1.24$ & 100 \\
MS 2254-37 & 22:57:39.0 & -36:06:07 & 0.039 & 1.15 & $1.84\pm0.43$ & 0.016 & 
$1530\pm120$ & $(9.49\pm4.22)\times10^{-2}$ & $0.143\pm0.022$ &
$5.50\pm1.49$ & 116 \\
RX J2304-51 & 23:04:39.0 & -51:27:59 & 0.106 & 1.33 & $3.2\pm0.2$ & 0.008 & 
$1775\pm130$ & $2.10\pm0.19$ & $1.36\pm0.07$ & $1.96\pm1.95$ & 114 \\
RX J2312-34 & 23:12:34.8 & -34:04:20 & 0.202 & 1.74 & $0.78\pm1.13$ & 0.017& 
$4200\pm950$ & $9.85\pm0.40$ & $6.62\pm0.15$ & $6.65\pm3.84$ & 119 \\
RX J2317-44 & 23:17:50.0 & -44:22:27 & 0.132 & 1.89 & $2.50\pm0.80$ & 0.010 & 
$1010\pm150$ & $6.24\pm0.34$ & $5.43\pm0.15$ & $0.69\pm1.03$ & 107 \\
RX J2340-53 & 23:40:23.0 & -53:28:57 & 0.321 & 1.18 & $2.1\pm0.6$ & 0.012 & 
$1565\pm80$ & $5.12\pm0.30$ & $3.32\pm0.01$ & $0.76\pm2.12$ & 114 \\
RX J2349-31 & 23:49:24.0 & -31:26:03 & 0.135 & 1.23 & $1.55\pm0.57$ & 0.010 & 
$4200\pm2000$ & $2.31\pm0.21$ & $2.07\pm0.09$ & $3.97\pm2.26$ & 74 \\
\enddata
\tablenotetext{a}{Taken from \citet{grup04a}, who do not list uncertainties.}
\tablecomments{Blank lines indicate an additional GALEX observation of the
object on the preceding line.
UV fluxes were extracted from GALEX plates using the 
IRAF {\it phot}
package.  Equatorial coordinates and ${\rm H}\beta$ FWHM come from 
\citet{grup04a} where possible and from \citet{grup99} otherwise.
Color excesses come from Simbad, and ${\rm N_{H}}$ values were 
extracted from \citet{dick90}.  X-ray spectral indices and count rates
were taken from \citet{grup01} wherever possible and from \citet{grup98}
otherwise.  All other parameters come from \citet{grup98}.
The names used here, which use the naming convention of \citet{grup98},
differ from the
object names included in the ROSAT All-Sky Catalog for the same objects.}
\end{deluxetable}

\clearpage
\begin{deluxetable}{cccc}
\tabletypesize{\scriptsize}
\tablewidth{0pc}
\tablecaption{Re-classified Broad-Line Seyfert Galaxies
\label{tabReclassified}}
\tablehead{\colhead{Object} & 
\multicolumn{2}{c}{FWHM(H$\beta$)[km ${\rm s}^{-1}$]} & 
\colhead {$\alpha_{\rm x}$} \\
\colhead{} & \colhead{Grupe et al. (1999)} & \colhead{Grupe et al. (2004)}
& \colhead{}}
\startdata
MS0117-28 & $2925\pm100$ & $1681\pm260$ & $2.51\pm0.66$ \\
RXJ0323-49 & $2075\pm250$ & $1680\pm250$ & $2.03\pm0.10$ \\
RXJ1117+65 & $2160\pm110$ & $1650\pm170$ & $2.50\pm0.49$ \\
RXJ2216-44 & $2200\pm130$ & $1630\pm130$ & $1.98\pm0.38$ \\
Mrk734 & $2230\pm140$ & --- & $2.0\pm0.2$ \\
\enddata
\end{deluxetable}

\clearpage
\begin{deluxetable}{llcccccc}
\tabletypesize{\scriptsize}
\tablewidth{0pc}
\tablecaption{Physical Correlations (Bentz Masses)
\label{tabBentz}}
\tablehead{
\multicolumn{2}{c}{Parameters} & 
\multicolumn{2}{c}{NLS1} & \multicolumn{2}{c}{BLS1} &
\multicolumn{2}{c}{Merged} \\
\colhead{} & \colhead{} & \colhead{$r_{S}$} & \colhead{Prob.} & 
\colhead{$r_{S}$} & \colhead{Prob.} & \colhead{$r_{S}$} & \colhead{Prob.}}
\startdata
$\nu F_{\nu}(1528{\rm \AA})/\nu F_{\nu}(2{\rm keV})$ & $M_{BH}$ & 0.47 & 0.10 & 0.20 & 
0.54 & 0.10 & 0.64 \\
$\nu F_{\nu}(1528{\rm \AA})/\nu F_{\nu}(2{\rm keV})$ & $L/L_{\rm edd}$ & 0.61 & 0.027 & 
0.36 & 0.55 & 0.36 & 0.054 \\
$\nu F_{\nu}(1528{\rm \AA})/\nu F_{\nu}(2{\rm keV})$ & $\nu L_{\nu}(1528{\rm \AA})$ & 0.92 & 
$2.2\times10^{-12}$ & 0.39 & 0.052 & 0.74 & $2.3\times10^{-10}$ \\
$\nu F_{\nu}(1528{\rm \AA})/\nu F_{\nu}(2{\rm keV})$ & $\nu L_{\nu}(5100{\rm \AA})$ & 0.82 & 
$7.2\times10^{-8}$ & 0.13 & 0.59 & 0.60 & $1.9\times10^{-6}$ \\
$\nu F_{\nu}(1528{\rm \AA})/\nu F_{\nu}(2271{\rm \AA})$ & $M_{BH}$ & 0.44 & 0.14 & 0.48 & 
0.052 & 0.59 & $6.6\times10^{-4}$ \\
$\nu F_{\nu}(1528{\rm \AA})/\nu F_{\nu}(2271{\rm \AA})$ & $L/L_{\rm edd}$ & 0.53 & 0.065 & 
0.34 & 0.54 & $-0.13$ & 0.56 \\
$\nu F_{\nu}(1528{\rm \AA})/\nu F_{\nu}(2271{\rm \AA})$ & $\nu L_{\nu}(1528{\rm \AA})$ & 0.52 & 
$3.5\times10^{-3}$ & 0.61 & $1.2\times10^{-3}$ & 0.48 & $2.4\times10^{-4}$ \\
$\nu F_{\nu}(1528{\rm \AA})/\nu F_{\nu}(2271{\rm \AA})$ & $\nu L_{\nu}(5100{\rm \AA})$ & 0.39 & 
0.037 & 0.37 & 0.067 & 0.35 & $9.7\times10^{-3}$ \\
$\alpha_{\rm x}$ & $M_{BH}$ & 0.48 & 0.10 & 0.05 & 0.84 & $-0.43$ & 0.019 \\
$\alpha_{\rm x}$ & $L/L_{\rm edd}$ & 0.27 & 0.50 & 0.14 & 0.63 & 0.53 & 
$2.7\times10^{-3}$ \\
$\alpha_{\rm x}$ & $\nu L_{\nu}(1528{\rm \AA})$ & 0.55 & $1.9\times10^{-3}$ & 0.12 & 
0.62 & 0.31 & 0.022 \\
$\alpha_{\rm x}$ & $\nu L_{\nu}(5100{\rm \AA})$ & 0.49 & $7.0\times10^{-3}$ & 0.12 & 
0.61 & 0.28 & 0.038 \\
FWHM(H$\beta$) & $M_{BH}$ & 0.83 & $4.7\times10^{-4}$ & 0.64 & $5.4\times10^{-3}$ & 0.90 & 
$1.0\times10^{-11}$ \\
FWHM(H$\beta$) & $L/L_{\rm edd}$ & $-0.54$ & 0.027 & $-0.20$ & 0.57 & $-0.80$ &
$8.8\times10^{-8}$ \\
FWHM(H$\beta$) & $\nu L_{\nu}(1528{\rm \AA})$ & 0.32 & 0.094 & 0.03 & 0.88 & 0.045 & 
0.75 \\
FWHM(H$\beta$) & $\nu L_{\nu}(5100{\rm \AA})$ & 0.36 & 0.055 & $-0.03$ & 0.88 & 
0.08 & 0.59 \\
FWHM(H$\beta$) & $M_{BH}/\dot{m}$ & 0.62 & 0.023 & 0.84 &
$2.2\times10^{-5}$ & 0.94 & $1.0\times10^{-14}$ \\
$M_{BH}$ & $L/L_{\rm edd}$ & $-0.04$ & 0.89 & 0.14 & 0.63 & $-0.59$ & 
$5.8\times10^{-4}$ \\
$M_{BH}$ & $\nu L_{\nu}(5100{\rm \AA})$ & 0.90 & $2.8\times10^{-5}$ 
& 0.54 & 0.025 & 0.53 & $2.6\times10^{-3}$ \\
$L/L_{\rm edd}$ & $\nu L_{\nu}(5100{\rm \AA})$ & 0.044 & 0.89 & 0.14 
& 0.63 & $-0.59$ & $5.8\times10^{-4}$ \\
\enddata
\end{deluxetable}

\clearpage
\begin{deluxetable}{llrcrcrc}
\tabletypesize{\scriptsize}
\tablewidth{0pc}
\tablecaption{Physical Correlation Results
\label{tabPhysicalCorrelation}}
\tablehead{
\multicolumn{2}{c}{Parameters} & 
\multicolumn{2}{c}{NLS1} & \multicolumn{2}{c}{BLS1} &
\multicolumn{2}{c}{Merged} \\
\colhead{} & \colhead{} & \colhead{$r_{s}$} & \colhead{Prob.} & 
\colhead{$r_{s}$} & \colhead{Prob.} & \colhead{$r_{s}$} & \colhead{Prob.}}
\startdata
$\nu F_{\nu}(1528{\rm \AA})/\nu F_{\nu}(2{\rm keV})$ & $M_{BH}$ & 0.53 & 0.065 & 0.52 & 
0.03 & 0.28 & 0.60 \\
$\nu F_{\nu}(1528{\rm \AA})/\nu F_{\nu}(2{\rm keV})$ & $L/L_{\rm edd}$ & 0.55 & 0.050 & 
0.16 & 0.59 & 0.29 & 0.62 \\
$\nu F_{\nu}(1528{\rm \AA})/\nu F_{\nu}(2{\rm keV})$ & $\nu L_{\nu}(1528{\rm \AA})$ & 0.92 & 
$2.2\times10^{-12}$ & 0.39 & 0.052 & 0.74 & $2.3\times10^{-10}$ \\
$\nu F_{\nu}(1528{\rm \AA})/\nu F_{\nu}(2{\rm keV})$ & $\nu L_{\nu}(5100{\rm \AA})$ & 0.82 & 
$7.2\times10^{-8}$ & 0.13 & 0.59 & 0.60 & $1.9\times10^{-6}$ \\
$\nu F_{\nu}(1528{\rm \AA})/\nu F_{\nu}(2271{\rm \AA})$ & $M_{BH}$ & 0.47 & 0.10 & 0.26 & 
0.50 & 0.44 & 0.016 \\
$\nu F_{\nu}(1528{\rm \AA})/\nu F_{\nu}(2271{\rm \AA})$ & $L/L_{\rm edd}$ & 0.44 & 0.13 & 
0.57 & 0.017 & $-0.06$ & 0.74 \\
$\nu F_{\nu}(1528{\rm \AA})/\nu F_{\nu}(2271{\rm \AA})$ & $\nu L_{\nu}(1528{\rm \AA})$ & 0.52 & 
$3.5\times10^{-3}$ & 0.61 & $1.2\times10^{-3}$ & 0.48 & $2.4\times10^{-4}$ \\
$\nu F_{\nu}(1528{\rm \AA})/\nu F_{\nu}(2271{\rm \AA})$ & $\nu L_{\nu}(5100{\rm \AA})$ & 0.39 & 
0.037 & 0.37 & 0.067 & 0.35 & $9.7\times10^{-3}$ \\
$\alpha_{\rm x}$ & $M_{BH}$ & 0.51 & 0.079 & 0.28 & 0.50 & $-0.28$ & 0.60 \\
$\alpha_{\rm x}$ & $L/L_{\rm edd}$ & 0.21 & 0.55 & 0.14 & 0.62 & 0.49 & 
$5.8\times10^{-3}$ \\
$\alpha_{\rm x}$ & $\nu L_{\nu}(1528{\rm \AA})$ & 0.55 & $1.9\times10^{-3}$ & 0.12 & 
0.62 & 0.31 & 0.022 \\
$\alpha_{\rm x}$ & $\nu L_{\nu}(5100{\rm \AA})$ & 0.49 & $7.0\times10^{-3}$ & 0.12 & 
0.61 & 0.28 & 0.038 \\
FWHM(H$\beta$) & $M_{BH}$ & 0.82 & $6.5\times10^{-4}$ & 0.42 & 0.097 & 0.77 & 
$7.7\times10^{-7}$ \\
FWHM(H$\beta$) & $L/L_{\rm edd}$ & $-0.39$ & 0.53 & $-0.47$ & 0.060 & $-0.79$ &
$2.7\times10^{-7}$ \\
FWHM(H$\beta$) & $\nu L_{\nu}(1528{\rm \AA})$ & 0.32 & 0.094 & 0.03 & 0.88 & 0.045 & 
0.75 \\
FWHM(H$\beta$) & $\nu L_{\nu}(5100{\rm \AA})$ & 0.36 & 0.055 & $-0.03$ & 0.88 & 
0.08 & 0.59 \\
FWHM(H$\beta$) & $M_{BH}/\dot{m}_{\rm edd}$ & 0.87 & $1.1\times10^{-4}$ & 0.76 &
$3.6\times10^{-4}$ & 0.89 & $6.7\times10^{-11}$ \\
$M_{BH}$ & $L/L_{\rm edd}$ & $-0.14$ & 0.66 & 0.08 & 0.77 & $-0.53$ & 
$2.8\times10^{-3}$ \\
$M_{BH}$ & $\nu L_{\nu}\left(5100{\rm \AA}\right)$ & 0.92 & $7.6\times10^{-6}$ & 0.63 & $7.1\times10^{-3}$ & 0.63 & $2.1\times10^{-4}$ \\
$L/L_{\rm edd}$ & $\nu L_{\nu}\left(5100{\rm \AA}\right)$ & 0.05 & 0.87 & 0.63 & $6.3\times10^{-3}$ & 0.14 & 0.55 \\
\enddata
\end{deluxetable}

\clearpage
\begin{deluxetable}{llrcrcrc}
\tabletypesize{\scriptsize}
\tablewidth{0pc}
\tablecaption{Observable Correlation Results
\label{tabFluxCorrelation}}
\tablehead{
\multicolumn{2}{c}{Parameters} & 
\multicolumn{2}{c}{NLS1} & \multicolumn{2}{c}{BLS1} & 
\multicolumn{2}{c}{Merged} \\
\colhead{} & \colhead{} & \colhead{$r_{s}$} & \colhead{Prob.} & 
\colhead{$r_{s}$} & \colhead{Prob.} & \colhead{$r_{s}$} & \colhead{Prob.}}
\startdata
$\nu F_{\nu}(1528{\rm \AA})/\nu F_{\nu}(2{\rm keV})$ & $\alpha_{\rm x}$ & 0.90 & 
$7.6\times10^{-11}$ & 0.56 & $1.6\times10^{-3}$ & 0.53 & $3.8\times10^{-5}$ \\
$\nu F_{\nu}(1528{\rm \AA})/\nu F_{\nu}(2{\rm keV})$ & 
$\nu F_{\nu}(1528{\rm \AA})/\nu F_{\nu}(2271{\rm \AA})$ & 0.52 & $3.7\times10^{-3}$ & 0.23 &
 0.51 & 0.37 & $5.8\times10^{-3}$ \\
$\nu F_{\nu}(1528{\rm \AA})/\nu F_{\nu}(2{\rm keV})$ & FWHM(H$\beta$) & 0.34 & 0.076 & 
$-0.19$ & 0.51 & $-0.05$ & 0.71 \\
$\nu F_{\nu}(1528{\rm \AA})/\nu F_{\nu}(2271{\rm \AA})$ & $\alpha_{\rm x}$ & $-0.26$ & 0.55 &
0.06 & 0.78 & $-0.19$ & 0.57 \\
$\nu F_{\nu}(1528{\rm \AA})/\nu F_{\nu}(2271{\rm \AA})$ & FWHM(H$\beta$) & -0.41 & 0.029 &
$-0.03$ & 0.88 & $-0.03$ & 0.81 \\
$\alpha_{\rm x}$ & FWHM(H$\beta$) & 0.33 & 0.88 & $-0.14$ & 0.61 & $-0.53$ & 
$4.1\times10^{-5}$ \\
\enddata
\tablecomments{Here $r_{s}$ is the Spearman rank order correlation
coefficient, and Prob. is the probability for $r_{s}$ to
appear at random in two data sets drawn from two
uncorrelated variables.}
\end{deluxetable}

\clearpage
\begin{deluxetable}{lcccccccc}
\tabletypesize{\scriptsize}
\rotate
\tablecaption{Properties of Outliers
\label{tabOutliers}}
\tablehead{
\colhead{ } & \colhead{$z$} & \colhead{$\alpha_{\rm x}$} & 
\colhead{$\alpha_{\rm ox}$} &
\colhead{${\rm N}_{H}$} & \colhead{FWHM(H$\beta$)}&
\colhead{$\log(L_{\rm x})$} & 
\colhead{$\log\bigl(\lambda L_{\lambda}(1528{\rm \AA})\bigr)$} &
\colhead{$\log\bigl(\lambda L_{\lambda}(5100{\rm \AA})\bigr)$} \\
\colhead{} & \colhead{} & \colhead{} & \colhead{} & \colhead{${\rm cm}^{-2}$} &
\colhead{${\rm km\ s}^{-1}$} & \colhead{${\rm erg\ s}^{-1}$}
& \colhead{${\rm erg\ s}^{-1}$} & \colhead{${\rm erg\ s}^{-1}$}}
\startdata
RXJ0134-43 & 0.237 & 6.70 & $3\pm82$ & $1.6\times10^{20}$ & $1160\pm80$ & 41.3 & $45.44\pm0.02$ & 44.9 \\
RXJ0136-35 & 0.289 & 4.90 & $2.0\pm0.4$ & $5.6\times10^{20}$ & $1320\pm120$ & 43.9 & $45.26\pm0.02$ & 44.4 \\
RXJ0323-49 & 0.071 & 2.03 & $1.17\pm0.06$ & $1.72\times10^{20}$ & $1680\pm250$ & 44.0 & $43.16\pm0.05$ & 43.5 \\
RXJ0902-07 & 0.089 & 2.17 & $1.19\pm0.09$ & $3.31\times10^{20}$ & $1860\pm150$ & 44.1 & $43.80\pm0.03$ & 43.3 \\
IC3599 & 0.021 & 3.20 & $0.9\pm0.1$ & $3.77\times10^{20}$ & $635\pm100$ & 42.7 & $40.66\pm0.19$ & 42.6 \\
IRASF1239+33 & 0.044 & 2.02 & $1.21\pm0.06$ & $1.35\times10^{20}$ & $1640\pm250$ & 43.4 & $42.69\pm0.50$ & 43.3 \\
RXJ1413+70 & 0.107 & 1.40 & $0.6\pm0.1$ & $1.93\times10^{20}$ & $4400\pm1000$ & 44.2 & $42.77\pm0.13$ & 43.8 \\
RXJ2144-39 & 0.140 & 3.40 & $1.1\pm0.1$ & $4.89\times10^{20}$ & $1445\pm120$ & 43.5 & $43.81\pm0.06$ & 43.7 \\
MS2254-37 & 0.039 & 1.80 & $0.7\pm0.1$ & $1.15\times10^{20}$ & $1530\pm120$ & 43.5 & $41.50\pm0.19$ & 43.3 \\
\enddata
\tablecomments{$L_{\rm x}$ is measured in the ROSAT band, from 0.2-2.0 keV
\citep{grup04a}.  $L_{\lambda}(5100\ {\rm \AA})$, FWHM(H$\beta$), 
$\alpha_{\rm x}$ and redshift all come from \citet{grup04a} where possible
and from \citet{grup98} otherwise.}
\end{deluxetable}

\clearpage

\clearpage

\clearpage

\clearpage

\clearpage
\begin{figure}
\epsscale{1.0}
\plotone{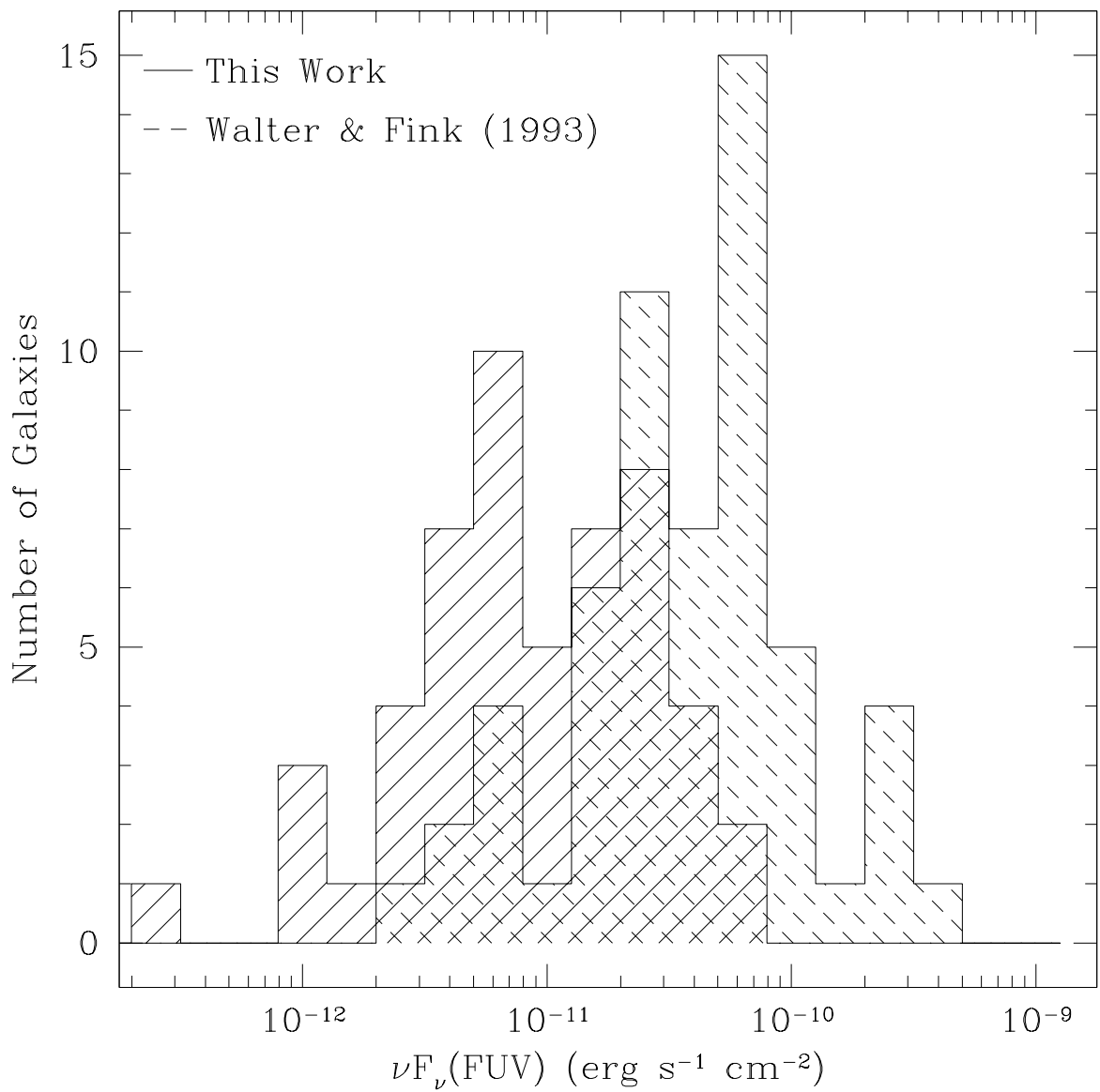}
\caption{Comparison of the far-UV fluxes used in the Walter \& Fink
(1993; WF93)
({\it solid fill}) analysis and those used in our analysis ({\it dashed fill}).
The two are not directly comparable, since
WF93 measure far-UV fluxes at 1375\AA\ while the
center of the GALEX $FUV$ band is at 1528\AA, but the comparison
is illuminating for to the analogy drawn in Fig. \ref{figSlope}
\label{figUvCompare}}
\end{figure}

\clearpage
\begin{figure}
\epsscale{1.0}
\plotone{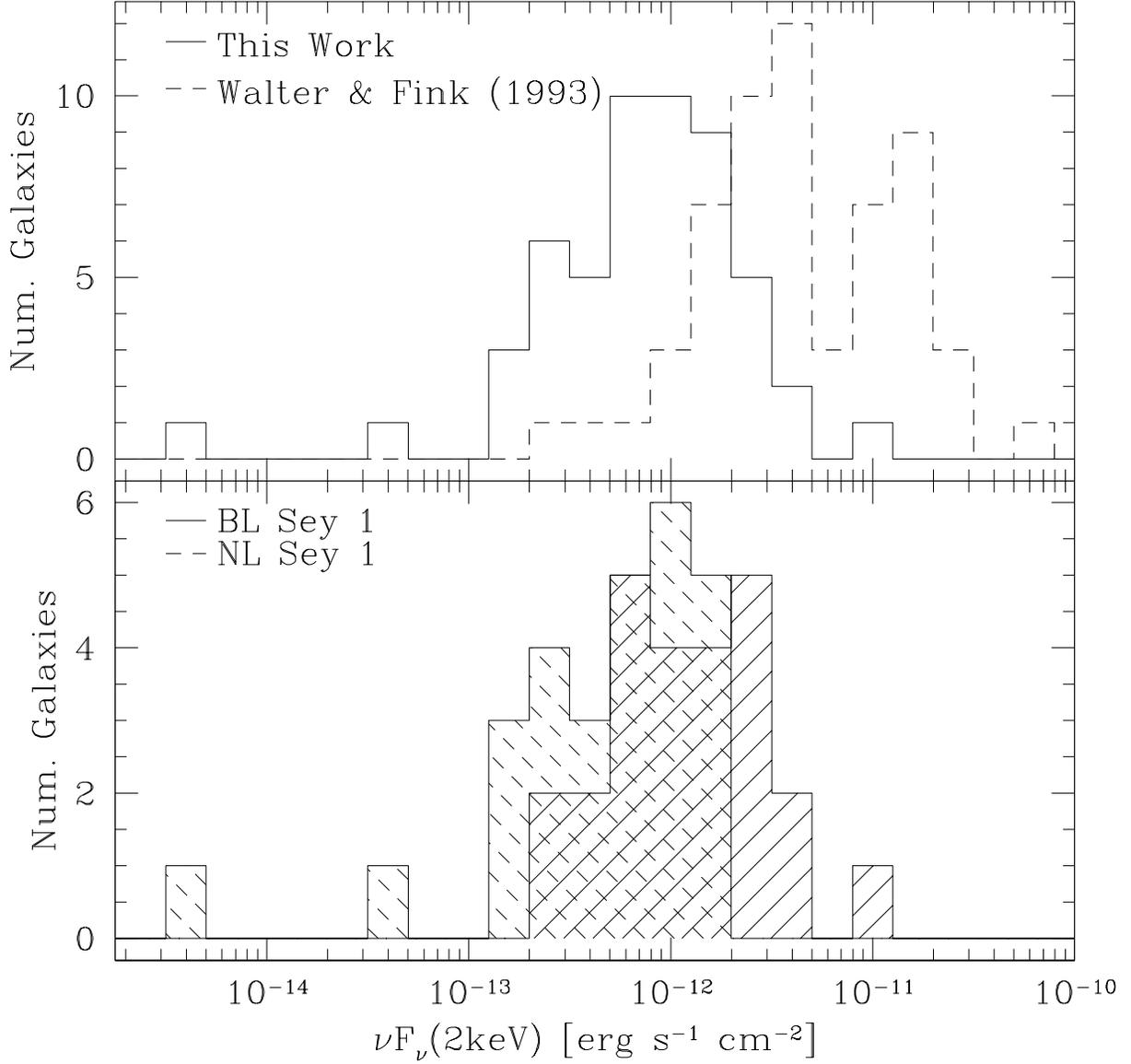}
\caption{Distributions of 2keV fluxes in our BLS1 ({\it solid fill}) and NLS1
({\it dashed fill}) samples (lower panel) and
a comparison of our combined sample ({\it solid}) with that of 
Walter \& Fink (1993; upper panel; {\it dashed}).  Our sample extends to
significantly lower $\nu F_{\nu}(2{\rm keV})$ than the Walter \& Fink sample
due to the steeper X-ray spectral indices in our objects.
\label{figXrayFlux}}
\end{figure}

\clearpage
\begin{figure}
\epsscale{1.0}
\plotone{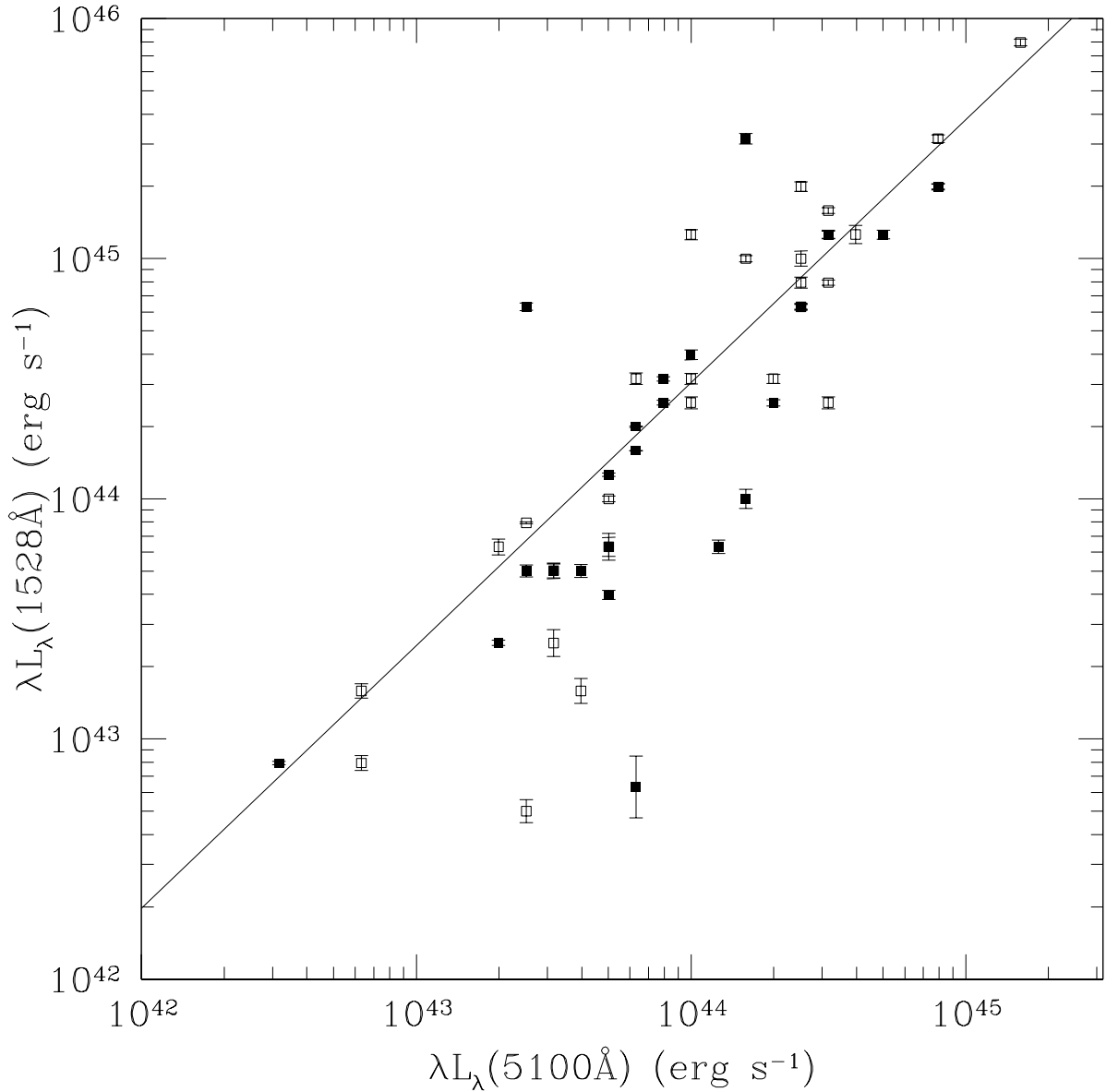}
\caption{Correlation of $FUV$ and $V$-band luminosities for both
BLS1 ({\it filled}) and NLS1 ({\it open}) points.  Objects with
unusually low $\lambda {\rm L}_{\lambda}(1528\ {\rm \AA})$ for their 
$\lambda {\rm L}_{\lambda}(5100\ {\rm \AA})$ could result from obscuration,
and abnormally large 
$\lambda {\rm L}_{\lambda}(1528\ {\rm \AA})$ might indicate star formation.
\label{figLuminosity}}
\end{figure}

\clearpage
\begin{figure}
\epsscale{1.0}
\plotone{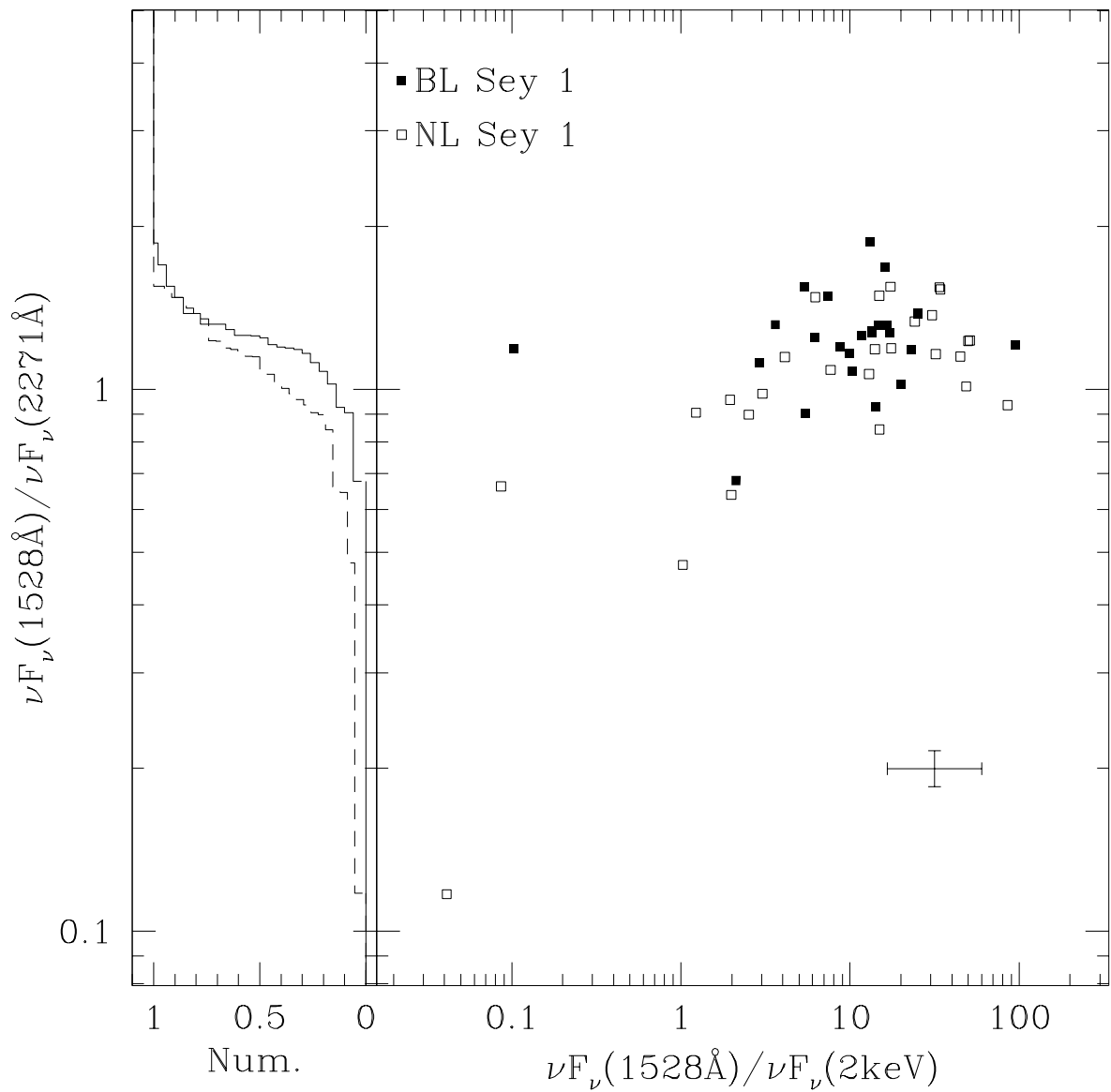}
\caption{Relation between the shape 
$\biggl(\nu F_{\nu}(1528\ {\rm \AA})/\nu F_{\nu}(2271\ {\rm \AA})\biggr)$ 
and strength 
$\biggl(\nu F_{\nu}(1528\ {\rm \AA})/\nu F_{\nu}(2\ {\rm keV})\biggr)$ 
of the Big Blue Bump.
The left panel shows the cumulative distributions of
$\nu F_{\nu}(1528\ {\rm \AA})/\nu F_{\nu}(2271\ {\rm \AA})$ in the BLS1 
({\it solid}) and NLS1 ({\it dashed}) samples.  The cross in the lower right
corner indicates typical error bars.
\label{figXrayShape}}
\end{figure}

\clearpage
\begin{figure}
\epsscale{1.0}
\plotone{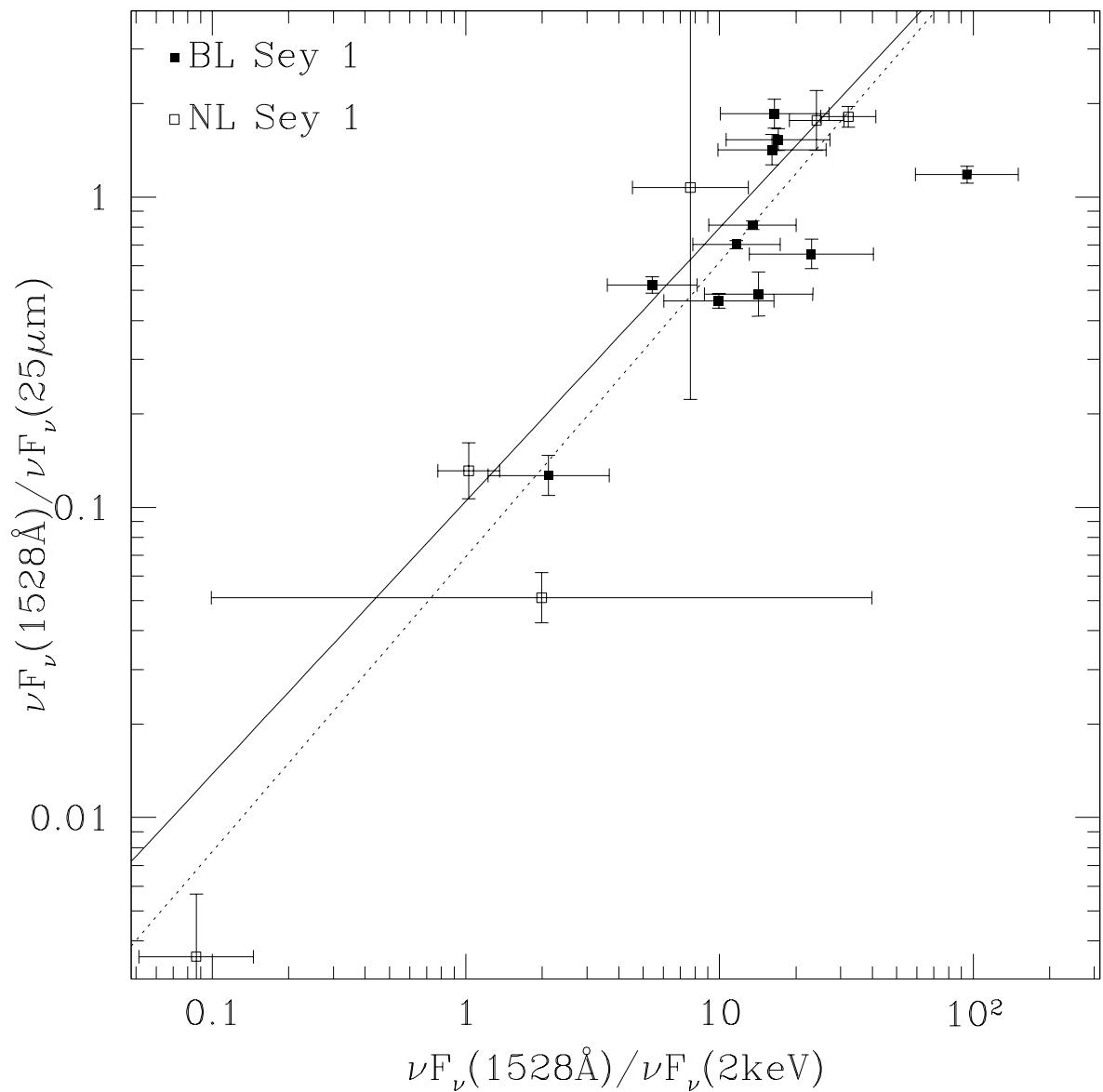}
\caption{Relation between two indicators for the strength of Big Blue Bump
with respect to the underlying, power-law continuum.
The {\it solid} line marks the average relation between the two ratios 
reported by \citet{walt93}, and the {\it dotted} line indicates the best
fit to our sample.
\label{figBumpStrength}}
\end{figure}

\clearpage
\begin{figure}
\epsscale{1.0}
\plotone{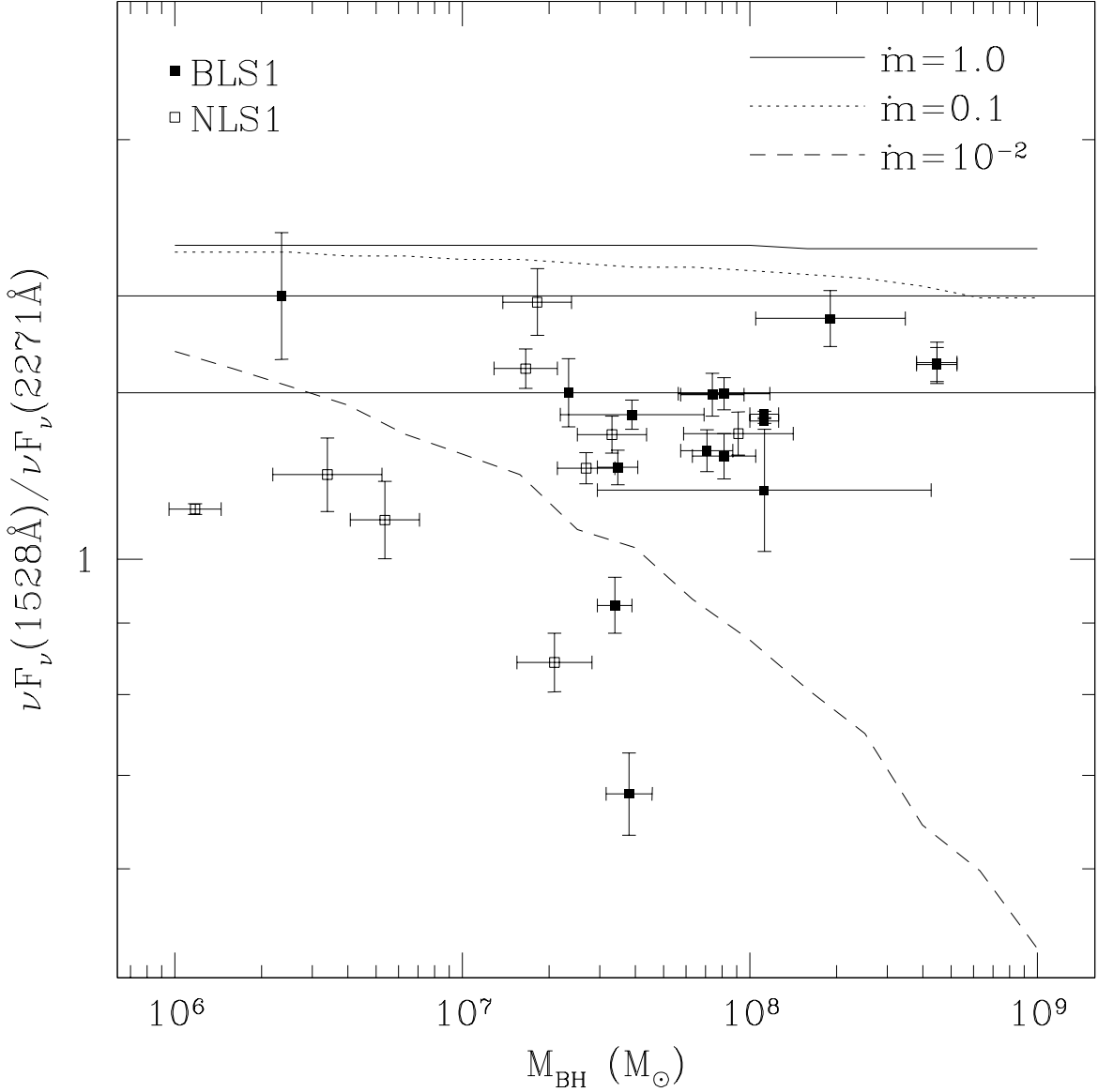}
\caption{Rest-frame GALEX flux ratios for a standard thin disk at various
black hole masses and Eddington ratios ($\dot{m}$) overplotted with flux
ratios and black hole masses for AGNs with measured $M_{\rm BH}$.  Model
ratios were determined by multiplying a multicolor blackbody disk from XSPEC
by the GALEX $FUV$ and $NUV$ bandpasses.  Eddington ratios were calculated
from the Grupe et al. bolometric luminosities, uncorrected for UV flux.
Outliers (Tab. \ref{tabOutliers}) are not shown.
The accretion rates
implied by these UV colors are significantly lower than the
measured values, suggesting there must be a contribution
from sources other than a standard thin disk.
Error bars include statistical errors only.
\label{figShapeMass}}
\end{figure}

\clearpage
\begin{figure}
\epsscale{1.0}
\plotone{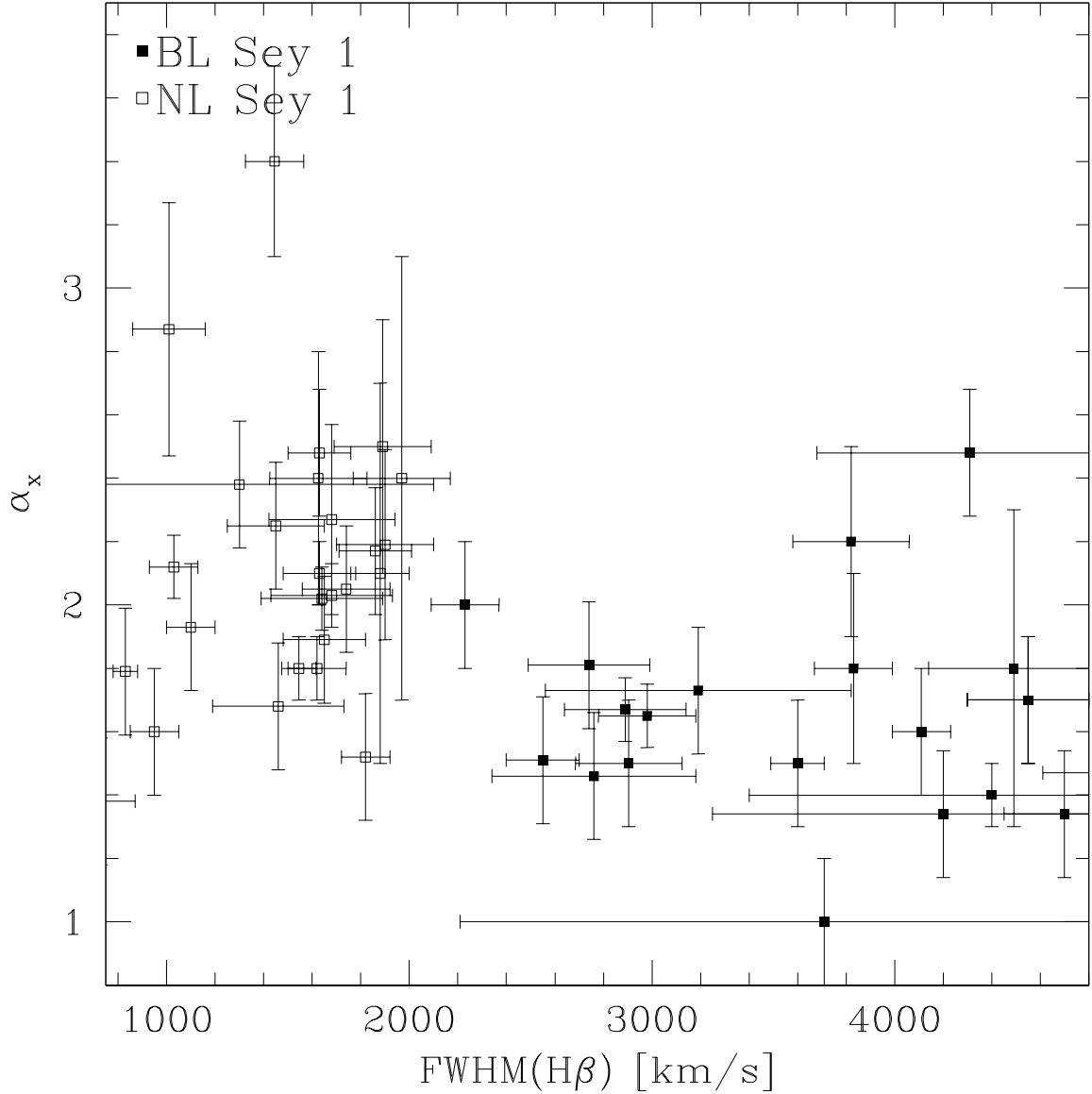}
\caption{Relations between FWHM(H$\beta$) and the shape
of the soft X-ray continuum.
The Spearman test indicates a correlation between these
parameters, but the data appear to be more
consistent with the ``zone of avoidance'' reported by
\citet{boll96}.
\label{figWidthCorr}}
\end{figure}

\clearpage
\begin{figure}
\epsscale{1.0}
\plotone{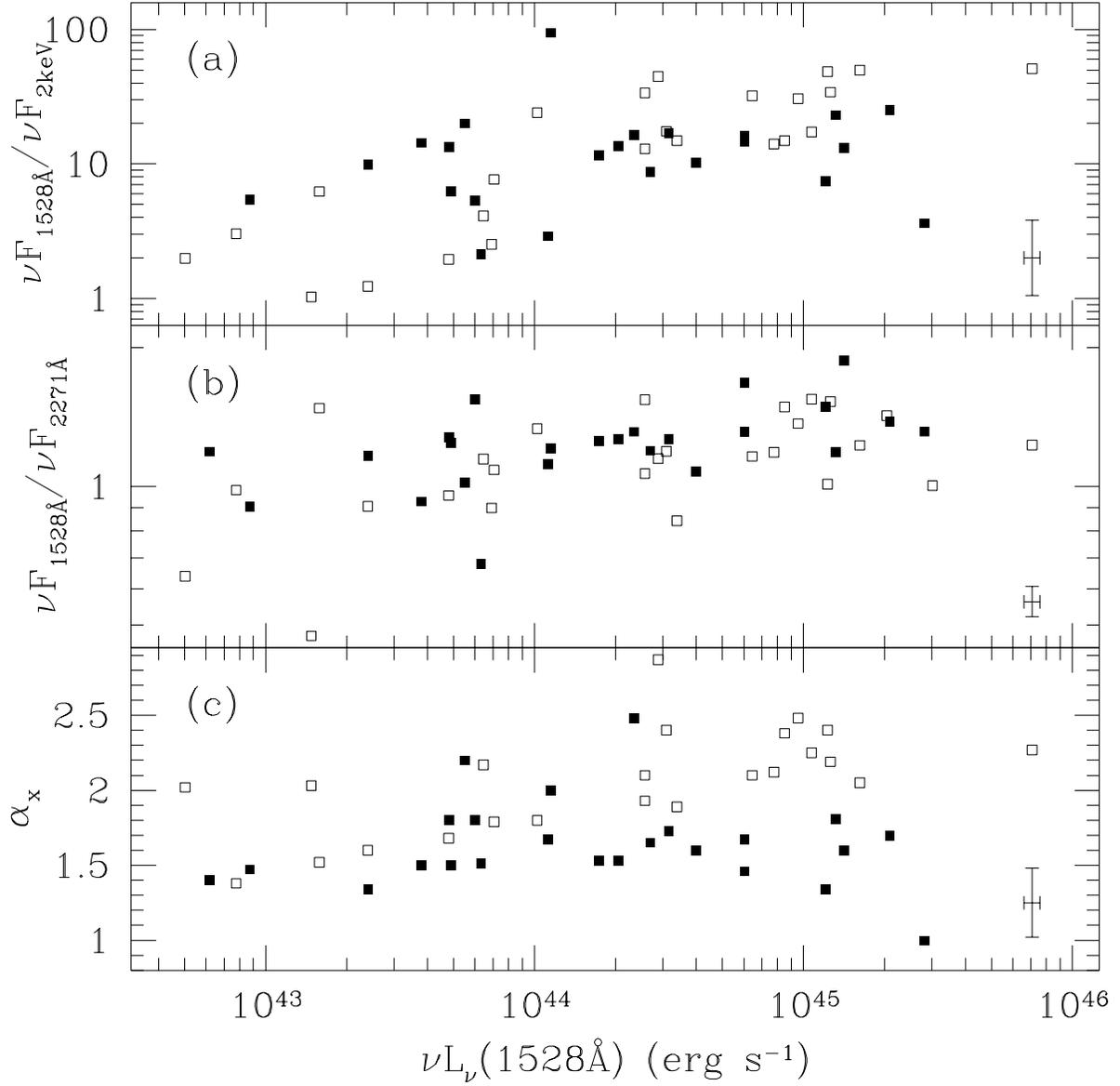}
\caption{Relations between $\nu L_{\nu}(1528\ {\rm \AA})$ and various observable
properties.  The trends are obviously stronger among the NLS1 ({\it open})
sample than among the BLS1 ({\it filled}) sample.  Error bars displayed on
each panel represent the average uncertainties after excluding the outliers.
\label{figLuminosityCorr}}
\end{figure}

\clearpage
\begin{figure}
\epsscale{1.0}
\plotone{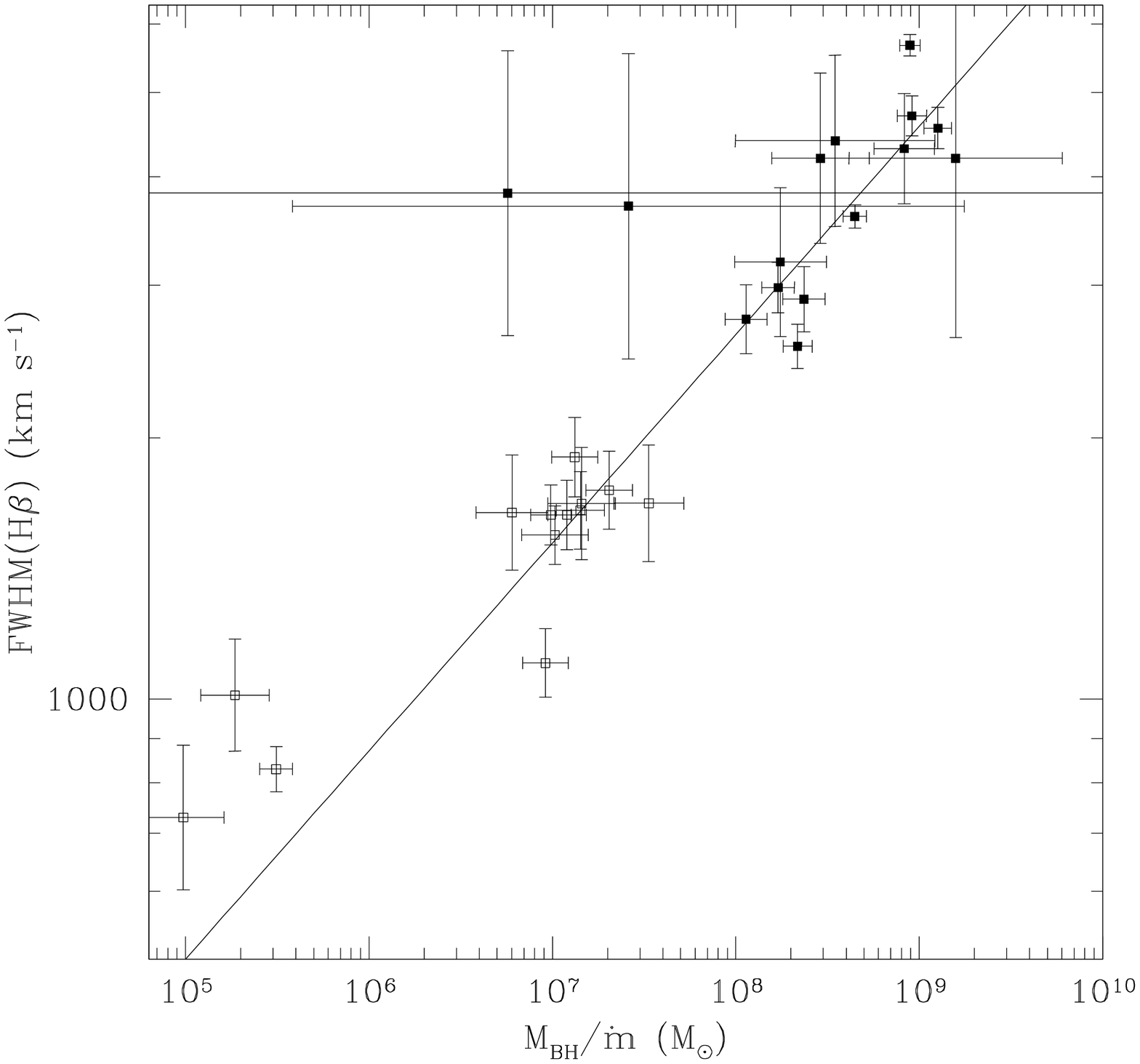}
\caption{Line width as a function of the ratio between 
black hole mass ($M_{BH}$) and 
Eddington ratio ($\dot{m}$).  Virial relationships predict 
${\rm w(H}\beta{\rm )}\propto (M_{BH}/\dot{m}_{\rm edd})^{1/4}$,
assuming that $R_{\rm BLR}\propto L^{1/2}$.
Black hole masses were determined using
the \citet{kasp00} relation, and Eddington ratios were calculated from
the bolometric luminosities listed in \citet{grup04a}.  The best fit
relation to the full sample is 
$\log[{\rm FWHM(H}\beta{\rm )}]=(0.24\pm0.02)\log[M_{BH}/(\dot{m}M_{\odot})]+(1.5\pm0.1)$,
which is consistent with a simple virialized structure for the broad line
region.  NLS1s
are shown with {\it open} points and BLS1s with {\it filled} points.
\label{figPhysicalWidth}}
\end{figure}

\clearpage
\begin{figure}
\epsscale{1.0}
\plotone{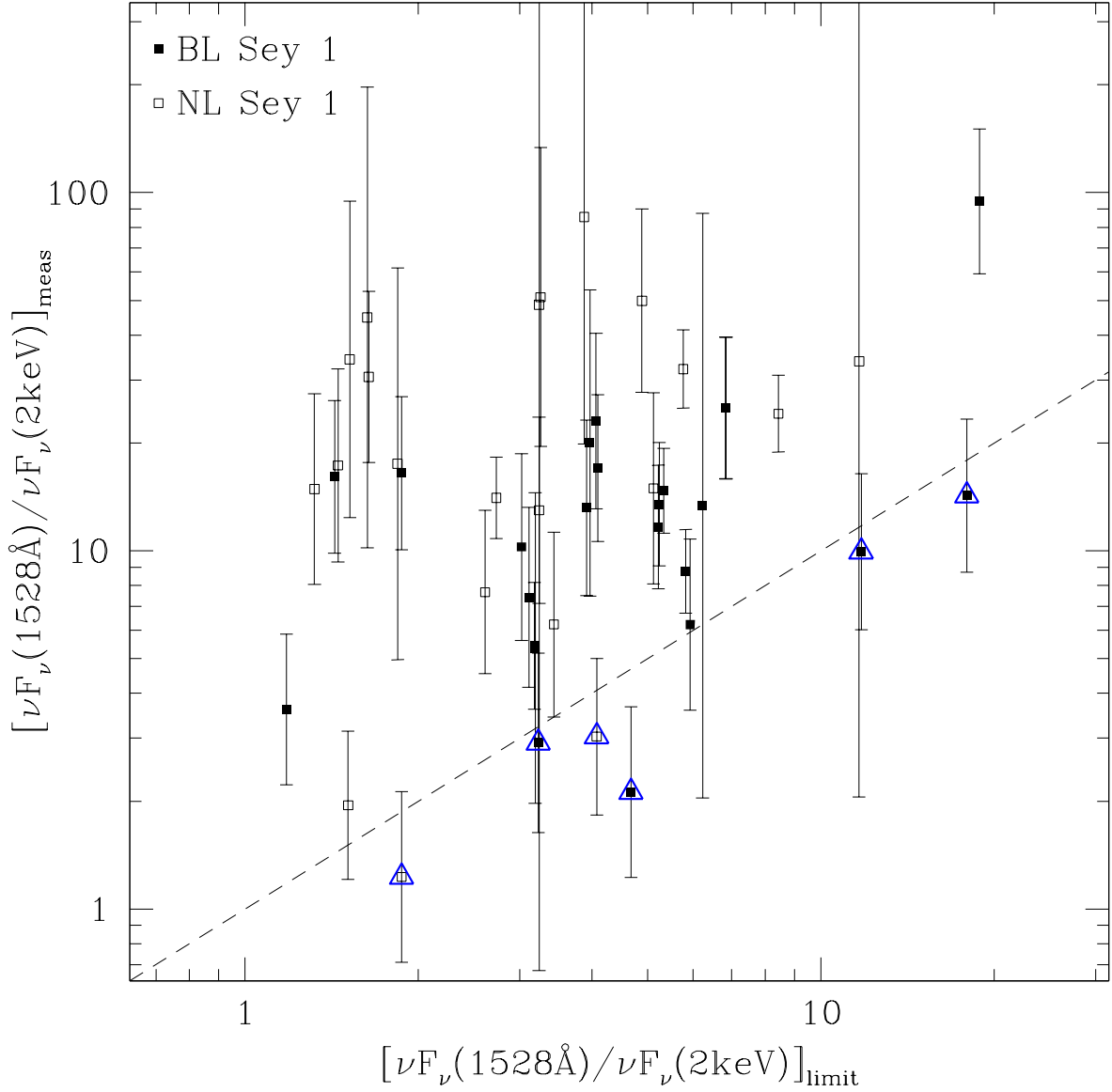}
\caption{Comparison of
$\nu F_{\nu}(1528\ {\rm \AA})/\nu F_{\nu}(2\ {\rm keV})$ lower limits
(x-axis) from our {\it ad hoc} model with the measured ratios (y-axis).
The {\it dashed line}
marks the line of equality, so any object to the right of the line
(emphasized by the {\it blue triangles}) have lower limits in excess of
the measured flux ratios.
\label{figRatioLimits}}
\end{figure}

\clearpage
\begin{figure}
\epsscale{1.0}
\plotone{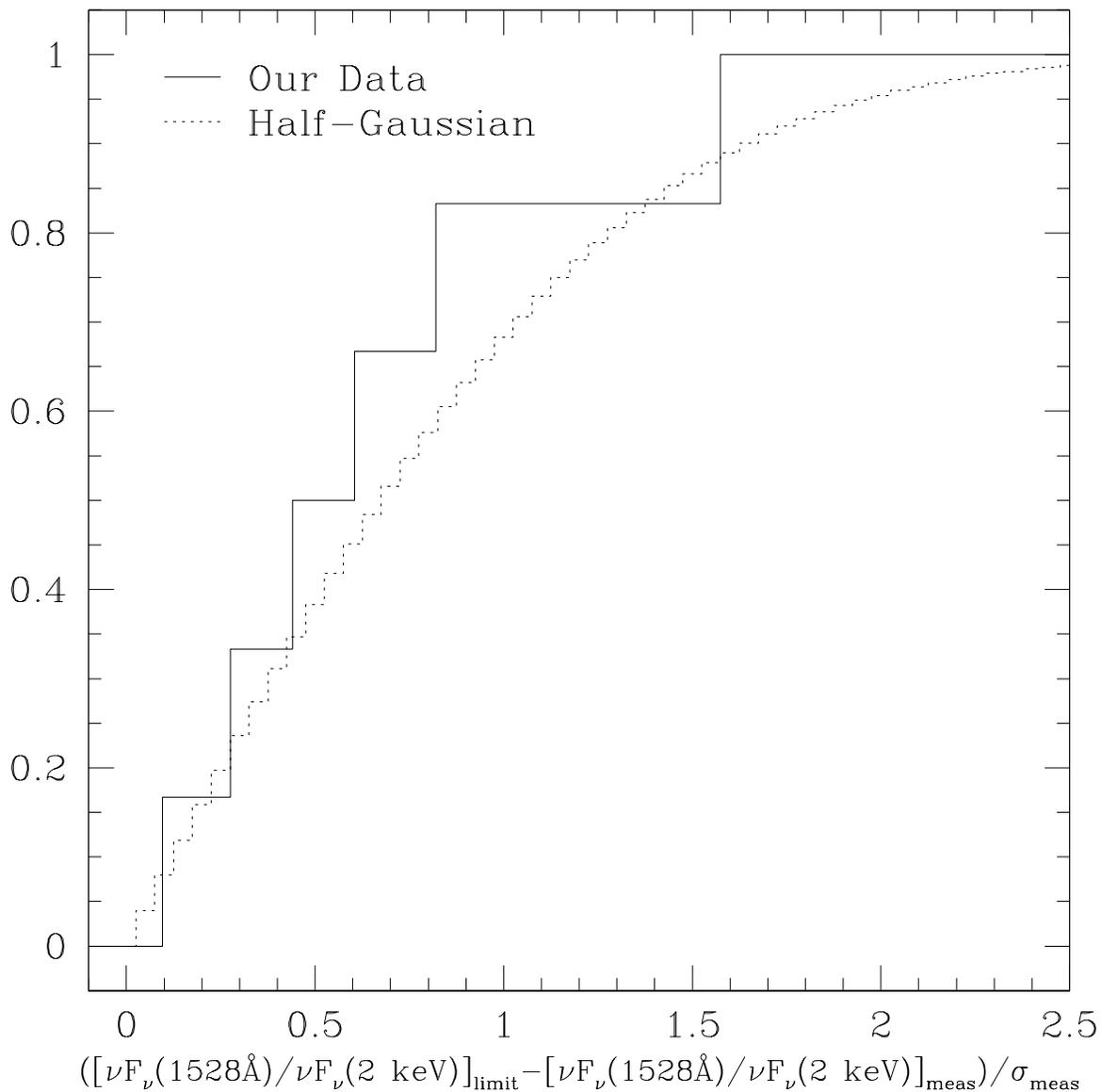}
\caption{Cumulative distribution function of objects whose lower limits
on $\nu F_{\nu}(1528\ {\rm \AA})/\nu F_{\nu}(2\ {\rm keV})$ exceed the
measured values ({\it solid}) compared with cumulative distribution 
of a half-Gaussian ({\it dotted}).
The half-Gaussian has distribution function
$c(x)=erf(x/\sqrt2)$.
\label{figLowerCum}}
\end{figure}


\begin{thebibliography}{}
\bibitem[Arnaud et al.(1985)]{arna85}Arnaud, M.A., et al. 1985, MNRAS, 217, 105
\bibitem[Bentz et al.(2006)]{bent06}Bentz, M., Peterson, B.M., Pogge, R.W., Vestergaard, M., \& Onken, C.A. 2006, \apj, 644, 133
\bibitem[Boller, Brandt \& Fink(1996)]{boll96}Boller, Th., Brandt, W.N. \& Fink, H. 1996, A\&A, 305, 53
\bibitem[Cardelli, Clayton \& Mathis(1989)]{card89}Cardelli, J.A., Clayton G.C., \& Mathis, J.S. 1989, \apj, 345, 245
\bibitem[Chen \& Wang(2004)]{chen04}Chen, L. \& Wang, J. 2040, \apj, 614, 101
\bibitem[Czerny \& Elvis(1987)]{czer87}Czerny, B. \& Elvis, M. 1987, \apj, 321, 305
\bibitem[Czerny et al.(2003)]{czer03}Czerny, B. et al. 2003, A\&A, 412, 317
\bibitem[Denney et al.(2009)]{denn09} Denney, K.~D., et al.\ 2009, arXiv:0904.0251
\bibitem[Dewangan et al.(2007)]{dewa07}Dewangan, G.C., Griffiths, R.C., Dasgupta, S. \& Rao, A.R. 2007, ApJ, 671, 1284
\bibitem[Dickey \& Lockman(1990)]{dick90}Dickey, J.M. \& Lockman, F.J. 1990, ARA\&A, 28, 215
\bibitem[Done \& Nayakshin(2007)]{done07} Done, C., \& Nayakshin, S.\ 2007, \mnras, 377, L59
\bibitem[Elvis et al.(1994)]{elvi94}Elvis, M. et al. 1994, ApJS, 95, 1
\bibitem[Gierli\'nski \& Done(2004)]{gier04}Gierli\'nski, M. \& Done, C. 2004, MNRAS, 349, 7
\bibitem[Gierli\'nski \& Done(2006)]{gier06}Gierli\'nski, M. \& Done, C. 2006, MNRAS, 371, 16
\bibitem[Grupe et al.(1998)]{grup98}Grupe, D., Beuermann, K., Thomas, H.-C., Mannheim, K., and Fink, H.H. 1998, A\&A, 330, 25
\bibitem[Grupe et al.(1999)]{grup99}Grupe, D., Beuermann, K., Mannheim, K., and Thomas, H.-C. 1999, A\&A, 350, 805
\bibitem[Grupe et al.(2001)]{grup01}Grupe, D., Thomas,H.-C. \& Beuermann, K. 2001, A\&A, 367, 470
\bibitem[Grupe(2004)]{grup04b}Grupe, D. 2004, \apj, 127, 1799
\bibitem[Grupe \& Mathur(2004)]{grup04c}Grupe, D. \& Mathur, S. 2004, \apj, 606, L41
\bibitem[Grupe et al.(2004)]{grup04a}Grupe, D., Wills, B.J., Leighly, K.M. \& Meusinger, H. 2004, \apj, 127, 156
\bibitem[Heinzeller, Mineshige \& Ohsuga(2006)]{hein06}Heinzeller, D., Mineshige, S. \& Ohsuga, K. 2006, MNRAS, 372, 1208
\bibitem[Heinzeller \& Duschl(2007)]{hein07}Heinzeller, D. \& Duschl, W.J. 2007, MNRAS, 374, 1146
\bibitem[Kaspi et al.(2000)]{kasp00}Kaspi, S., et al. 2000, \apj, 533, 631
\bibitem[Kawaguchi, Shimura \& Mineshige(2001)]{kawa01}Kawaguchi, T., Shimura, T. \& Mineshige, S. 2001, \apj, 546, 966
\bibitem[Kawaguchi(2003)]{kawa03}Kawaguchi, T. 2003, \apj, 593, 69
\bibitem[Kelly et al.(2008)]{kell08} Kelly, B.~C., Bechtold, J., Trump, J.~R., 
Vestergaard, M., \& Siemiginowska, A.\ 2008, \apjs, 176, 355
\bibitem[Kuraszkiewica et al.(2000)]{kura00}Kuraszkiewicz, J., Wilkes, B.J., Czerny, B. \& Mathur, S. 2000, \apj, 542, 692
\bibitem[Laor \& Netzer(1989)]{laor89}Laor, A. \& Netzer, H. 1989, MNRAS, 238, 897
\bibitem[Laor et al.(1994)]{laor94}Laor, A., Fiore, F., Elvis, M., Wilkes, B.J., McDowell, J.C. 1994, \apj, 435, 611
\bibitem[Leighly \& Moore(2004)]{leig04}Leighly, K.M. \& Moore, J.R. 2004, \apj, 611, 107
\bibitem[Mainieri et al.(2007)]{main07}Mainieri, V., et al. 2007, \apj, 172, 368
\bibitem[Mannheim, Schulte \& Rachen(1995)]{mann95}Mannheim, K., Schulte, M. \& Rachen, J. 1995, A\&A, 303, L41
\bibitem[Marconi et al.(2008)]{marc08}Marconi, A., Axon, D.~J., Maiolino, R.,
Nagao, T., Pastorini, G., Pietrini, P., Robinson, A., \& Torricelli, G.\ 2008,
\apj, 678, 693
\bibitem[Marconi et al.(2009)]{marc09}Marconi, A., Axon, D., Maiolino, R., 
Nagao, T., Pietrini, P., Risaliti, G., Robinson, A., \& Torricelli, G.\ 2009, 
arXiv:0905.0539
\bibitem[McHardy et al.(2006)]{mcha06}McHardy, I.M., Koerding, E., Knigge, C., Uttley, P. \& Fender, R.P. 2006, Nature, 444, L730
\bibitem[Morrissey et al.(2005)]{morr05}Morrissey, P., et al. 2005, \apj, 619, 7
\bibitem[Muchotrzeb \& Paczy\'nski(1982)]{much82}Muchotrzeb, B. \& Paczy\'nski, B. 1982, Acta Astron., 32, 1
\bibitem[Netzer(2009)]{netz09}Netzer, H. 2009, \apj, 695, 793
\bibitem[Nied{\'z}wiecki \& Zdziarski(2006)]{nied06}Nied{\'z}wiecki, A., \& Zdziarski, A.~A.\ 2006, \mnras, 365, 606
\bibitem[Onken et al.(2004)]{onke04}Onken, C.A., et al. 2004, \apj, 615, 645
\bibitem[Peterson et al.(2000)]{pete00}Peterson, B.M., et al. 2000, 542, 161
\bibitem[Piconcelli et al.(2005)]{pico05}Piconcelli, E., Jimenez-Bail\'on, E.,
Guainazzi, M., Schartel, N., Rodr\'iguez-Pascual, P.M., Santos-Lle\'o, M. 2005,
A\&A, 432, 15
\bibitem[Piro et al.(1997)]{piro97}Piro, L., Matt, G., \& Ricci, R.\ 1997, \aaps, 126, 525
\bibitem[Pounds, Done \& Osborne(1995)]{poun95}Pounds, K.A., Done, C. \& Osborne, J.P. 1995, MNRAS, 277, L5
\bibitem[Puchnarewicz et al.(1995a)]{puch95a}Puchnarewicz, E.M., Mason, K.O., Siemiginowska, A. \& Pounds, K.A. 1995, MNRAS, 276, 20
\bibitem[Puchnarewicz et al.(1995b)]{puch95b}Puchnarewicz, E.M., Granduardi-Raymont, G., Mason, K.O. \& Sekiguchi, K. 1995, MNRAS, 276, 1281
\bibitem[Ross \& Fabian(1993)]{ross93}Ross, R.R. \& Fabian, A.C. 1993, MNRAS, 261, 74
\bibitem[R\'o\.za\'nska et al.(2002)]{roza02}R\'o\.za\'nska, A., Dumont, A.-M., Czerny, B., \& Collin, S. 2002, MNRAS, 332, 799
\bibitem[Salim et al.(2007)]{sali07}Salim, S. et al. 2007, ApJS, 173, 267
\bibitem[Schlegel et al.(1998)]{schl98}Schlegel, D.J., Finkbeiner, D.P. \& Davis, M. 1998, \apj, 500, 525
\bibitem[Schurch \& Done(2006)]{schu06}Schurch, N.J. \& Done, C. 2006, MNRAS, 371, 81
\bibitem[Schurch \& Done(2007)]{schu07}Schurch, N.J. \& Done, C. 2007, MNRAS, 381, 1413
\bibitem[Scott et al.(2004)]{scot04}Scott, J.E. et al. 2004, \apj, 615, 135
\bibitem[Shakura \& Sunyaev(1973)]{shak73}Shakura, N.I. \& Sunyaev, R.A. 1973, A\&A, 24, 337
\bibitem[Sobolewska \& Done(2007)]{sobo07} Sobolewska, M.~A., \& Done, C.\ 2007, \mnras, 374, 150
\bibitem[Strateva et al.(2005)]{stra05}Strateva, I.V., Brandt, W.N., Schneider, D.P., Vanden Berk, D.G. \& Vignali, C. 2005, \apj, 130, 387
\bibitem[Turner \& Pounds(1989)]{turn89}Turner, T.J. \& Pounds, K.A. 1989, MNRAS, 240, 833
\bibitem[Walter \& Fink(1993)]{walt93}Walter, R. \& Fink, H.H. 1993, A\&A, 274, 105
\bibitem[Walter et al.(1994)]{walt94}Walter, R., et al. 1994, A\&A, 285, 119
\bibitem[Wang \& Netzer(2003)]{wang03}Wang, J.-M. \& Netzer, H. 2003, A\&A, 398, 927
\bibitem[Watson, Mathur \& Grupe(2007)]{wats07}Watson, L.C., Mathur, S. \& Grupe, D. 2007, \apj, 133, 2435
\end{thebibliography}
\end{document}